\documentclass[twocolumn,english,showpacs,nofootinbib]{revtex4-1}
\usepackage[T1]{fontenc}
\usepackage{geometry}
\usepackage{xcolor}
\usepackage{babel}
\usepackage{units}
\usepackage{amsmath}
\usepackage{amssymb}
\usepackage{stmaryrd}
\usepackage{graphicx}
\usepackage{wasysym}
\usepackage[colorlinks=true,citecolor=magenta,linkcolor =gray]{hyperref}
\usepackage[all]{hypcap} 

\makeatletter

\IfFileExists{lmodern.sty}{\usepackage{lmodern}}{}

\makeatother

\begin{document}
\title{Observation of trap-assisted formation of atom-ion bound states}

\author{Meirav Pinkas}

\affiliation{Department of Physics of Complex Systems and AMOS, Weizmann Institute of Science,
Rehovot 7610001, Israel}

\author{Or Katz}
\thanks{Present address: Duke Quantum Center, Department of Physics and Electrical and Computer Engineering, Duke University, Durham, NC 27701}
\affiliation{Department of Physics of Complex Systems and AMOS, Weizmann Institute of Science, Rehovot 7610001, Israel}

\author{Jonathan Wengrowicz}
\affiliation{Department of Physics of Complex Systems and AMOS, Weizmann Institute of Science,
Rehovot 7610001, Israel}

\author{Nitzan Akerman}
\affiliation{Department of Physics of Complex Systems and AMOS, Weizmann Institute of Science,
Rehovot 7610001, Israel}

\author{Roee Ozeri}
\affiliation{Department of Physics of Complex Systems and AMOS, Weizmann Institute of Science,
Rehovot 7610001, Israel}

\begin{abstract}

Pairs of free particles cannot form bound states in an elastic collision due to momentum and energy conservation. In many ultracold experiments, however, the particles collide in the presence of an external trapping potential which can couple the center-of-mass and relative motions and assist the formation of bound states. Here, we report on observation of weakly bound molecular states formed between one ultracold $^{87}$Rb atom and a single trapped $^{88}$Sr$^+$ ion in the presence of a linear Paul trap. We show that bound states can form efficiently in binary collisions, and enhance the rate of inelastic processes. By observing electronic spin-exchange rate, we study the dependence of these bound states on the collision energy and magnetic field and extract the average molecular binding energy $E_{\textrm{bind}}=0.7(1)$ mK$\cdot k_B$ and the mean lifetime of the molecule $\tau=0.5(1)\,\mu$s, with good agreement with molecular-dynamics simulations. Our simulations predict a power-law distribution of molecular lifetimes with a mean that is dominated by extreme, long-lived, events. The dependence of the molecular properties on the trapping parameters opens new avenues to study and control ultracold collisions. 

\end{abstract}
\maketitle
\section*{introduction}
Collisions between pairs of particles are among the fundamental building blocks of molecular formation and quantum chemistry. Owing to energy and momentum conservation, pairs of free particles cannot bind in binary elastic collisions; instead, formation of molecules usually requires inelastic dynamics or three-body interactions, as realized in processes such as photo-association \cite{Jones2006,idziaszek2011multichannel,tomza2019cold,jyothi2016photodissociation}, Feshbach-association \cite{Kohler2006,Chin2010,hirzler2022observation,drews2017inelastic,weckesser2021observation} or three body recombination \cite{harter2012single,mohammadi2021life,krukow2016energy}.

In many ultracold collision experiments, however, the particles are not free but rather trapped by an external potential, such as optical-dipole traps \cite{Grimm2000,Anderegg2019,Lambrecht2017,Reynolds2020,schmidt2020optical} or ion traps \cite{Grier2009,Ratschbacher2012,Hall2012,Meir2016,Saito2017,Joger2017,weckesser2021observation,mohammadi2021life}. Charged particles, in particular, are highly susceptible to electric fields and usually require strong electromagnetic potentials to assist the trapping. In most trapped ion experiments, the ions are held in a Paul trap, using time-dependent electric fields \cite{Paul1990,Leibfried2003}.

Various studies have shown that the presence of a trap can modify the properties of collisions and lead to emergence of unique phases \cite{Hadzibabic2006,Kinoshita2004}, to change in the profile of scattering resonances \cite{Olshanii1998,Mies2000,Kinoshita2004,Haller2010,Kestner2010}, to non-equilibrium dynamics \cite{Rouse2017,Meir2018} or to formation of bound states via adiabatic merge of two different traps \cite{Stock2003,idziaszek2007controlled,Melezhik2016,melezhik2019impact}.
Yet, to date, trap-assisted bound states between pairs of neutral atoms and atomic ions have not been observed. 

Here we show that the confinement of the ion can lead to formation of a short-lived and loosely bound $^{88}$Sr$^+$-$^{87}$Rb molecule in a cold binary collision. 
By measuring the probability of electronic spin-exchange (SE) between the atom and ion at different settings, we estimate the molecule's lifetime and the binding energy. We compare our results to a molecular dynamics (MD) simulation and characterize the effect of the trap for various experimental configurations. We show that the lifetime of these molecules is power-law distributed such that the mean molecular lifetime is dominated by extreme, long-lived, events.

\section*{Formation and detection of atom-ion bound states}
In most hybrid atom-ion experiments in the ultracold regime, both neutral atom and ion are trapped by external potentials. Yet, the dynamics is predominantly governed by the ion's trapping potential owing to its considerably higher trapping frequency. This potential breaks the translation invariance symmetry of the interaction Hamiltonian and could hence generate coupling between the relative motion and the center-of-mass motion of the atom-ion pair. This coupling could transfer energy between these two frames and reduce the pair's energy in the relative frame, and by that, allow for formation of short-lived bound states. A similar mechanism, inelastic confinement-induced resonances, leads to a creation of a bound state of ultracold atoms in an anharmonic trap \cite{Sala2012,Sala2013,Lee2022SpinDynamics,Capecchi2022Observation} or when the particles have different trapping frequencies \cite{Melezhik2009}.

\begin{figure*}
    \centering
    \includegraphics{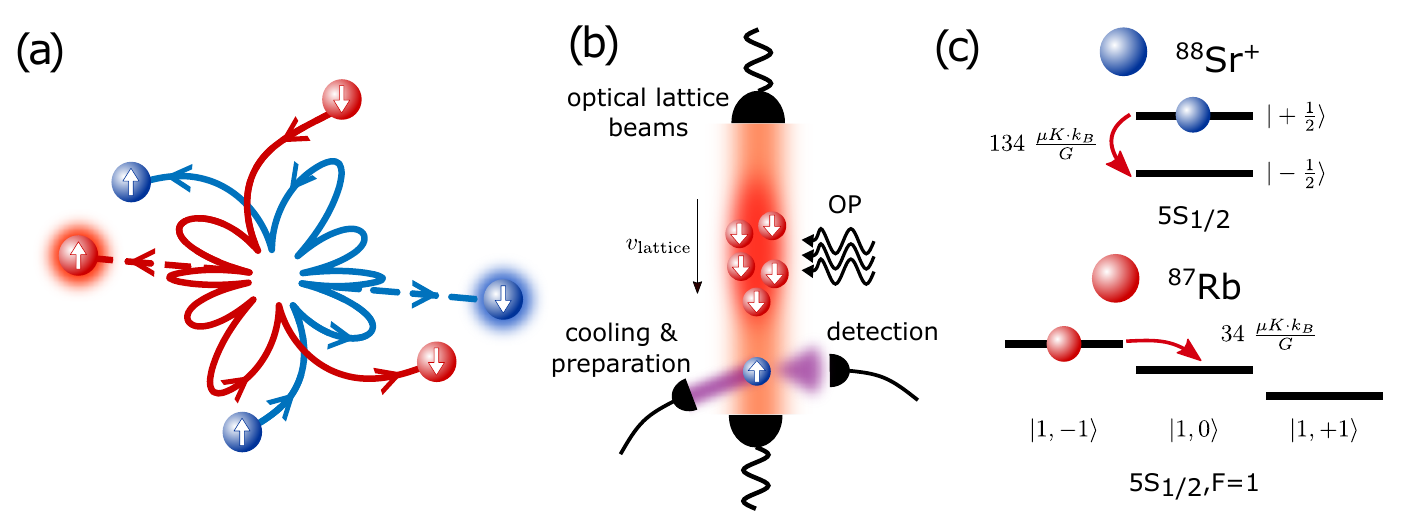}
    \caption{\textbf{Bound state formation and experimental apparatus.} (a) A collision between an atom and an ion in center-of-mass frame when the ion is trapped in a harmonic trap. A bound state with five short-range collisions is formed (solid lines). If an exothermic spin exchange occurs, shown here schematically in the second collision, the energy is released and the bound state may dissociate (dashed lines). (b) The experimental setup. A cloud of Rb atoms is trapped, cooled down, and loaded into an optical lattice trap. The atoms are optically pumped (OP) into a specific spin state. The atomic cloud is transported over the ion at a velocity $v_\textrm{lattice}$. The ion is prepared in a specific spin state prior to the passage of the cloud, and typically a single Langevin collision occurs. Detection of the ion state is realized afterwards using state-selective fluorescence. (c) Energy levels for $^{88}$Sr$^+$ and $^{87}$Rb in the electronic ground state. After a spin-exchange process, denoted by the red arrows for an exothermic process, the Zeeman energy is released or absorbed to the motional degrees of freedom.}
    \label{fig:bound_sim_energy_scheme}
\end{figure*}

We illustrate the formation of a bound state in Fig.~\ref{fig:bound_sim_energy_scheme}(a), via numerical simulation of a collision in the presence of an ideal, spherical symmetric harmonic potential that applies trapping forces only on the ion (see Methods). Here the atom and ion bind in a binary collision; Their relative motion manifests multiple oscillations, shown in the center-of-mass frame. Consequently, during the lifetime of this molecule, the pair comes into close contact multiple times and by that enhances the action of short-range chemical forces during the scattering, which increases the probability of inelastic processes. 

The total inelastic scattering probability of an ion inside a cloud of neutral atoms depends on the number of Langevin scattering events with different atoms and on the inelastic process probability in a single Langevin scattering event with one atom. The number of distinct scattering events with different atoms is determined by the Langevin capture rate coefficient, $K_L$, which depends on the mass of the atoms and the neutral atom polarizability \cite{tomza2019cold}. In each Langevin collision, the atom and ion spiral towards each other, and have a single period of interaction if they remain free, or multiple periods of short-range interactions if a bound state is formed. Owing to the short-range of the chemical forces that lead to inelastic scattering, the probability of inelastic processes is primarily determined by the relative molecular potential curves at a scale of a few Bohr radii \cite{Cote2000,tomza2019cold}. On the other hand, there are two characteristic length scales that are associated with the external trap. The first is the harmonic oscillator length, $r_\textrm{ho}\sim \sqrt{\hbar/m\omega}$, where $m$ is the ion mass and $\omega$ is the trapping frequency. The second is the distance at which the polarization potential is equal to the ion’s trapping potential $r\sim \sqrt[6]{C_4/m\omega^2}$ \cite{Cetina2012}. Both length scales are about a few tens of nanometers and are considerably larger than the range of the chemical forces. Owing to this length scale separation, the probability of inelastic short-range is unaffected by the trapping potential. Yet, the formation of bound states can lead to multiple short-range encounters, and by that, to the enhancement of inelastic processes upon dissociation.

\section*{experimental system}
We study the formation of loosely bound $^{87}$Rb-$^{88}$Sr$^+$ molecules using the experimental system that is shown in Fig.~\ref{fig:bound_sim_energy_scheme}(b) and detailed in Refs.~\cite{Katz2022,lattice2021,Meir2018}. In brief, we trap and cool $^{87}$Rb atoms in a magneto-optical trap (MOT) with a dark-MOT stage followed by a polarization gradient cooling. A cloud of $(5-10)\times10^5$ atoms is loaded into a 1D optical lattice created by two counter-propagating Nd:YAG laser beams \cite{lattice2021}. The atoms are optically pumped into one of the hyperfine states $|F=1,m_F=\pm1\rangle$ in the electronic ground-state. The atomic cloud is transported through the Paul trap at a fixed velocity of 0.14 mm/ms, which is controlled by the relative frequency difference between the optical lattice beams.
A $^{88}$Sr$^+$ ion is trapped in a linear segmented Paul trap and is cooled using Doppler cooling, followed by resolved-sideband cooling into the ground state ($\bar{n}\lesssim0.5$ for each of the motional modes). Then, the ion is optically pumped into one of the two Zeeman states of the electronic ground-state manifold $5S_{1/2}$. Following preparation, and before overlapping with the atomic cloud, the magnetic field is changed from 3 G to a target value between 0.5 G and 20 G using a pair of Helmholtz coils. Excess-micromotion (EMM) is compensated every hour during the experiment to less than 50 $\mu$K for each of the radial modes and to a few $\mu$K in the axial direction.

After the atomic cloud passes the ion, the magnetic field is ramped back to its initial value for state-detection of the ion. Due to the long relaxation time of the magnetic field in the system, the detection takes place $70$ ms after the current in the quantization coils returns to its initial value. We measure the spin state of the ion by double-shelving of one of the ground state levels into two different Zeeman states in the 4D$_{5/2}$ manifolds using a narrow linewidth laser at 674 nm, followed by detection of state-dependent fluorescence light using the strong 5S$_{1/2}$-5P$_{1/2}$ dipole transition.

\section*{Enhancement of spin-exchange rate}
 Spin-exchange is one of the most probable ineslastic processes that occurs in collisions of $^{88}$Sr$^+$ and $^{87}$Rb \cite{Cote2000,Aymar2011}, which can be enhanced by the formation of  molecules. 
 In Fig.~\ref{fig:bound_sim_energy_scheme}(c) we illustrate a spin exchange process between $^{88}$Sr$^+$ and $^{87}$Rb in their lower hyperfine manifolds. In the presence of a magnetic field, the exchange of spin is accompanied by exchange of internal magnetic energy and kinetic energy. We studied the exothermic channel via measurement of spin-flip of the ion when the pair is prepared in the initial state $|\uparrow\rangle_{Sr^+}|1,-1\rangle_{Rb}$, and also the endothermic channel when the initial state is $|\downarrow\rangle_{Sr^+}|1,+1\rangle_{Rb}$. The energy gap for these two pathways depends on the magnetic field and equal to $\Delta_\textrm{SE}=\pm h\cdot$3.5 $\frac{\textrm{MHz}}{\textrm{G}}=\pm$168 $\frac{\mu\textrm{K}\cdot k_B}{\textrm{G}}$, where $h$ is Planck's constant and $k_B$ is Boltzmann's constant.

In the experiment, the mean number of Langevin capturing events per passage of the atomic cloud through the ion trap is given by $N_L = \rho K_L/v_\textrm{lattice}$ (see Methods and Ref. \cite{lattice2021}), where $\rho$ is the atomic column density of the neutral atoms integrated along the lattice direction of motion, and $v_\textrm{lattice}$ is the speed of the lattice. The Langevin capture rate coefficient, $K_L$, is nearly independent of the energy of the colliding pair or the magnetic field and is given by $K_L=\pi\sqrt{4C_4/\mu}$ \cite{tomza2019cold}, where $C_4$ is the leading long-range induction coefficient of $^{87}$Rb, and $\mu$ is the reduced mass. 
All experiments are taken at a constant lattice velocity, and therefore the probability for a cold Langevin collision is constant, independent of the collision energy, and equal to 0.32(3) (Methods). Due to the Poisson distribution of the Langevin collisions, the probability for a single Langevin collision is 0.26, and the probability for multiple collisions is 0.06 (see Methods). The Langevin probability calibration process is described in Ref. \cite{Katz2022suppression} and in the Methods section. 

\subsection{Energy dependence}

\begin{figure}
    \centering

    \includegraphics[width=1\linewidth]{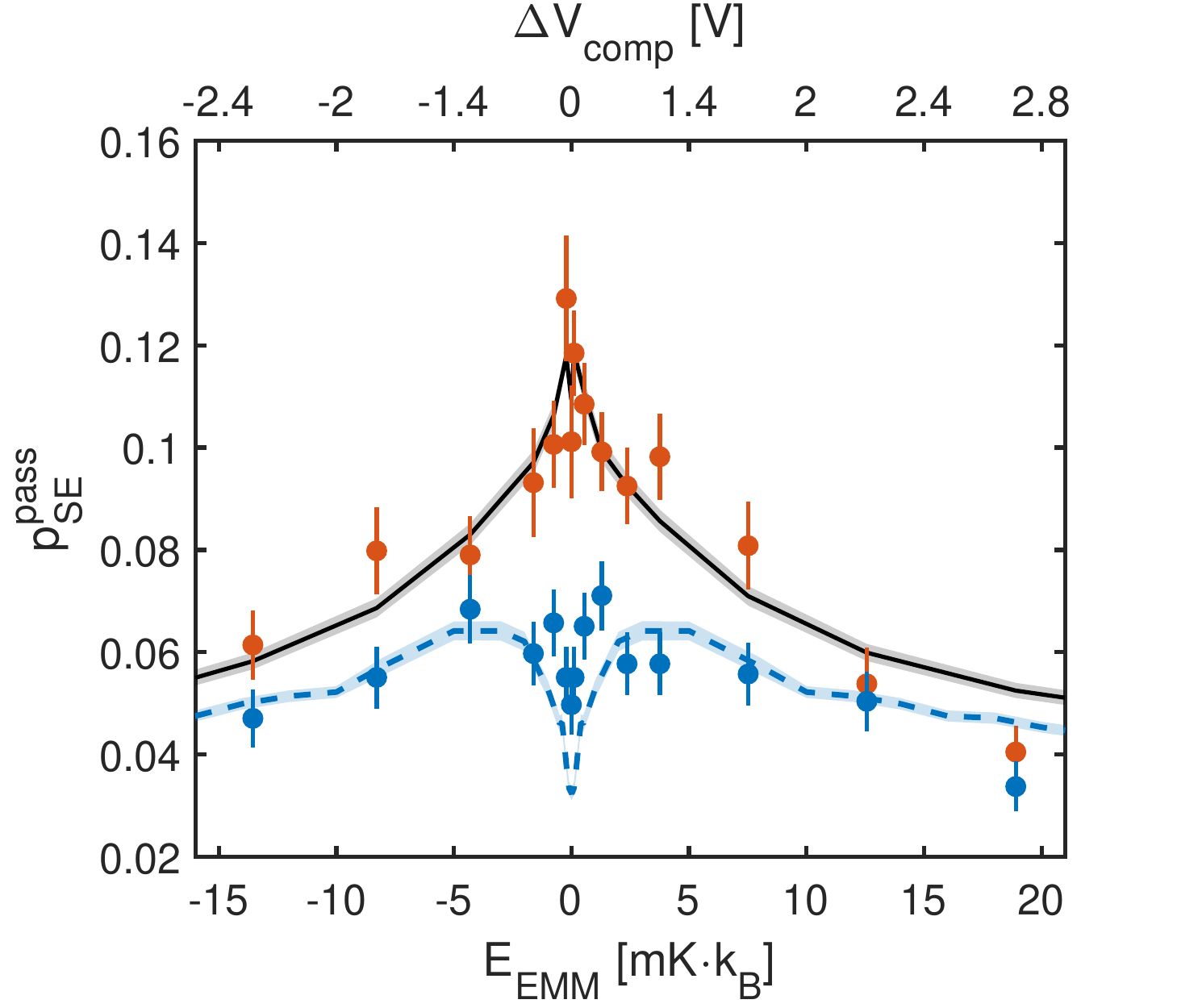}
    \caption{\textbf{Spin-exchange as a function of collision energy.} Spin exchange probability per passage as function of excess-micromotion energy is measured at 16 G for exothermic (red) and endothermic (blue) channels. Error bars are binomial distribution standard deviation. Black solid line is a maximum likelihood fit of to the short-range spin-exchange probability based on a numerical simulation, including detection efficiency change due to heating, giving a probability of $p^\textrm{0}_\textrm{SE}=0.122(4)$. Shaded area is 1$\sigma$ confidence bounds of the fit. Dashed blue line is the effective spin-exchange given by the simulation for the endothermic at B = 16 G transition when the short-range probability is 0.12, including detection efficiency.}
    \label{fig:SE_EMM}
\end{figure}

We first study the dependence of spin-exchange on the ions' excess micromotion (EMM) energy $E_{\textrm{EMM}}$, which is the energy in the fast oscillatory motion in the frequency of the trapping RF fields. $E_{\textrm{EMM}}$ is determined by the voltage difference $\Delta V_{\textrm{comp}}$, which is the difference between the voltage applied on one of the trap electrodes and the compensated voltage. In Fig.~\ref{fig:SE_EMM} we show the dependence of the measured spin-exchange probability per passage, $p^\textrm{pass}_{\textrm{SE}}$, of the cloud on the micromotion energy at a background field of $B=16$ G. 

While in the absence of a bound state we expect $p^\textrm{pass}_{\textrm{SE}}$ to be independent of the micromotion energy, our measurement shows enhancement of the exothermic channel at small $E_{\textrm{EMM}}$.
The measured ratio of spin-exchange probability and the probability of a Langevin collision varies between $\sim$10$\%$ at high EMM to $\sim$30$\%$ at low EMM. 
The enhancement of spin-exchange probability at low micromotion energies is an indication of the formation of a bound state, which leads to multiple short-range collisions.

For the endothermic channel, at low micromotion energy and for $B=16$ G, the spin-exchange probability is suppressed in comparison with the exothermic channel at the same energy. The exchange probability of the endothermic channel should be suppressed with respect to the exothermic one, if the kinetic energy of the bodies or the work done by the trap is smaller than the magnetic energy gap. At 16 G, the energy gap for an endothermic transition is about 2.5 mK, well above the collision energy at low EMM values. The lack of complete suppression of endothermic spin-exchange is another indication of trap associated effects, that could be explained by multiple short-range collisions during which work can be done by the trap time-dependent fields. 

\subsection{Magnetic field dependence}

\begin{figure}
    \centering
    \includegraphics[width=1\linewidth]{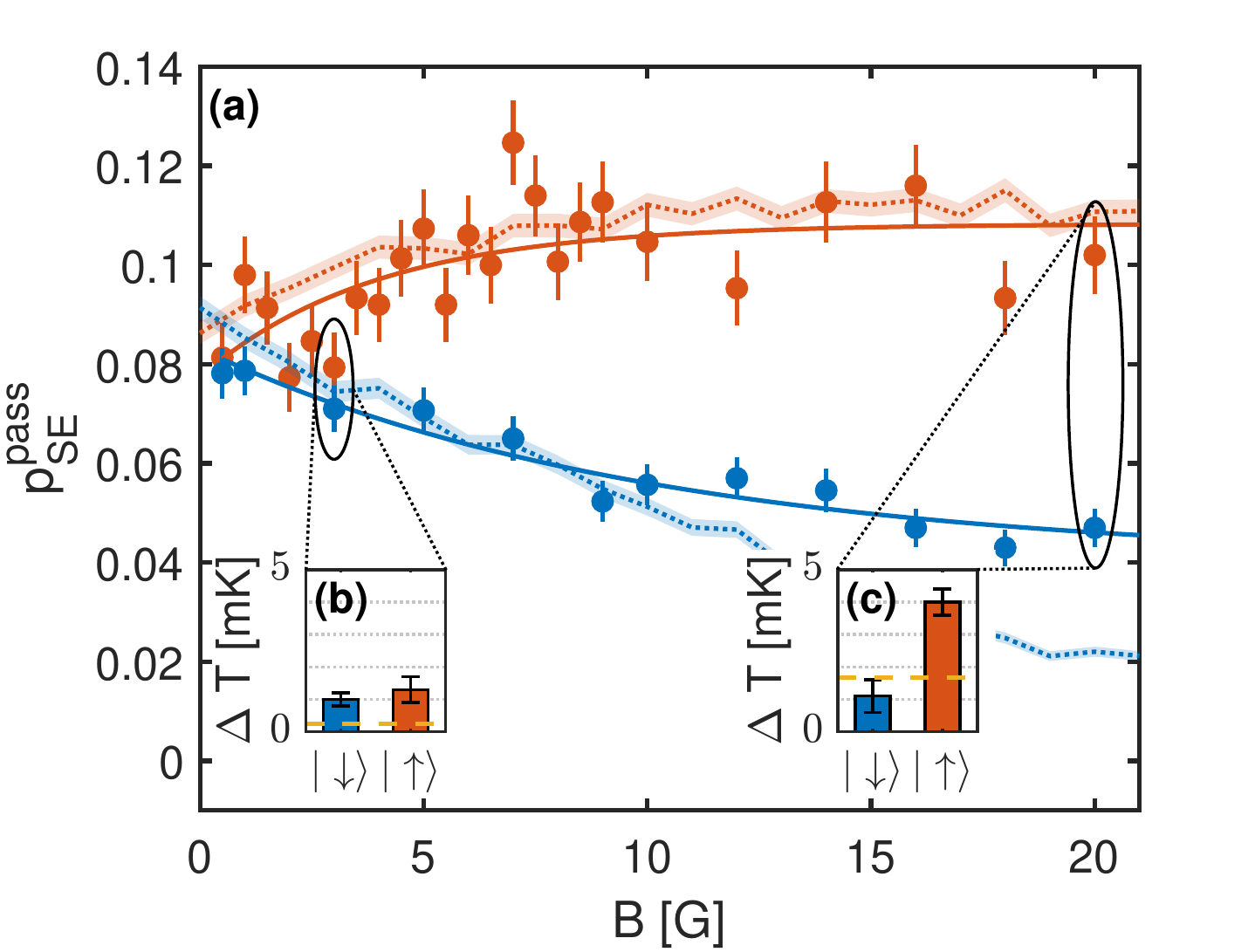}
    \caption{\textbf{Spin-exchange as a function of the magnetic field and thermometry.} (a) Spin-exchange probability per passage as a function of the magnetic field for the exothermic (red) and endothermic (blue) channels at lattice velocity 0.14 mm/ms (equivalent to 100$\mu$K). Error bars are binomial distribution standard deviation.
    Solid lines are maximum-likelihood estimation fit to a decay exponential function $p^\textrm{pass}_{\textrm{SE}}=A \cdot\exp(-B/B_0)+C$ for both experiments. The parameters are, A=-0.031(4), B$_0$=3.9(7) G and C=0.108(2) for the exothermic channel, and A=0.043(3), B$_0$=10(1) G and C=0.040(1) for the endothermic channel. Dotted lines are expected $p^\textrm{pass}_\textrm{SE}$ for the exothermic (red) and endothermic channels (blue) from the MD simulation, assuming the short-range $p^\textrm{0}_\textrm{SE}=0.12$ that was found using the data in Fig.~\ref{fig:SE_EMM} with no fit parameters. (bc) Heating of the ion using Rabi carrier thermometry on the spin-exchange event of the two transitions at 3 G (b) and 20 G (c). The dashed yellow line denotes the energy of half of the Zeeman gap.}
    \label{fig:SE_B}
    \end{figure}

We next turned to measuring spin-exchange dynamics at various magnetic fields. Owing to the conversion between internal magnetic energy and kinetic energy, 
we expect to see enhancement or suppression of spin-exchange as a function of the magnetic field for different scattering pathways if a bound state is formed. The measured endothermic (exothermic) spin-exchange probability as a function of the magnetic field is shown by the blue (red) filled circles in Fig.~\ref{fig:SE_B}. 

The dependence of the exothermic channel on the magnetic field can be understood from its effect on the bound state's dissociation. At low magnetic fields, reversal of the ion's spin to its original state in subsequent short-range encounters with the atom during the lifetime of the bound state acts to decrease the measured spin-exchange probability of the ion spin. In contrast, at magnetic fields higher than the binding energy, the bound state dissociates following the energy release of the first spin-exchange event, therefore suppressing further reversal of the spin, resulting in a higher spin-exchange probability.

For the endothermic channel, spin-exchange still occurs in the highest measured magnetic field of $20$ G, despite the fact it corresponds to an energy gap of above $3$ mK, which is larger than all other initial energy scales in the system. We further characterized the dynamics by measuring the temperature of the ion, post-selected on experiments in which its spin had flipped. The measured temperatures are shown in Fig.~\ref{fig:SE_B}(b-c), using the Rabi carrier thermometry technique \cite{Meir2016}. A reference temperature of non spin-exchange events was measured separately for each channel (see Methods). At a magnetic field of $3$ G, both exothermic (red) and endothermic (blue) pathways leave the ion at a similar temperature, of about $1$ mK, which is likely dominated by heating driven by the fields of the Paul trap \cite{Cetina2012,Meir2016}. At the highest magnetic field of 20 G, after an exothermic transition, the ion heats up to a temperature comparable with the Zeeman gap, of about $4$ mK. Surprisingly, after an endothermic transition, the temperature of the ion is similar to that measured at the low magnetic field, without any observed reduction of the kinetic energy, which is naively expected because of the Zeeman energy barrier. This might indicate additional trap-induced effects in this regime.

\section*{Molecular Dynamics Simulations}

\begin{figure*}
    \centering
    \includegraphics[width=1\textwidth]{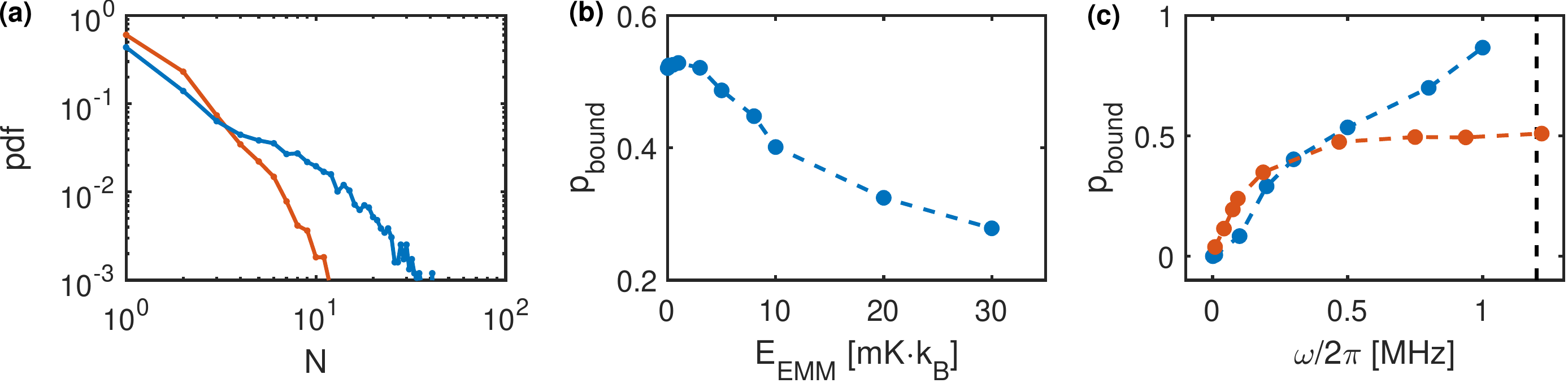}
    \caption{\textbf{Bound state properties from a molecular dynamics (MD) simulation.} (a) Probability density function (pdf) of the number of short-range collisions between Sr$^+$-Rb pair when the EMM is compensated (blue), and equivalent to 10 mK$\cdot k_B$ (red). (b) Bound state formation probability, in a single Langevin collision, as a function of EMM energy, for the experimental parameters. (c) Bound states formation probability, in a single Langevin collision in a harmonic trap (blue) and a Paul trap (red) as a function of trap frequency. The harmonic trap is a spherical symmetric trap with a frequency $\omega$. The Paul trap is a linear trap with an axial trapping frequency $\omega_{ax}=2\pi\cdot 3$ kHz, RF frequency $\Omega_{RF}=2\pi\cdot 26.51$ MHz of the drive and secular radial trapping frequency $\omega$, for an ion with zero initial energy. Black dashed line denotes the mean radial frequency in the experiment. In all graphs, the magnetic field is zero, and the Rb atoms have a velocity equivalent to 100 $\mu$K. Binomial confidence bounds of 1$\sigma$ are smaller than markers size.}
    \label{fig:Bound_sim}
\end{figure*}

We compared our observations to a molecular-dynamics (MD) model, numerically simulating the dynamics of the collisions in a Paul trap (see Methods). Using the simulation, we can study the distribution of number short-range collisions, $N$. We find that for our experimental trapping parameters, a bound state is typically formed in a Langevin collision when excess micromotion is compensated ($E_{\textrm{EMM}}=0$), as can be seen as blue line in Fig.~\ref{fig:Bound_sim}(a). In addition, the tail of the distribution exhibits a power-law behavior, which scales roughly as $1/N^2$. For this power-law distribution, the mean number of short-range collisions, and also the mean molecular lifetime, is dominated by rare tail events. 

When the collision energy is increased beyond the bound state's binding energy, the probability for more than one short-range collision, $p_\textrm{bound}$, is reduced, as can be seen in Fig.~\ref{fig:Bound_sim}(b). This coincides with the observed reduced spin-exchange probability at larger EMM in Fig.~\ref{fig:SE_EMM}. However, bound states still occur with high probability up to EMM energies of 10 mK$\cdot k_{\textrm{B}}$. When the collision occurs at high EMM, the collision energy depends on the momentary phase of the RF field, and therefore low energy events are still possible, leading to the formation of bound states. In addition, the power-law tail of the distribution is suppressed at higher values of micromotion, e.g.~at $E_{\textrm{EMM}}=10$ mK$\cdot k_{\textrm{B}}$ (red line in Fig. \ref{fig:Bound_sim}(a)). 

From the simulation we extracted the numerical probability distribution function (PDF) of the number of short-range collisions, $N$, for each micromotion energy. We then calculated the effective spin-exchange probability per a single passage for the high magnetic field exothermic process (see Methods). 
Then, by a maximum likelihood estimation to the measured data, we determined the microscopic spin-exchange probability per one period of short-range interaction in the exothermic channel of $p^0_\textrm{SE}=0.122(4)$. This estimation of microscopic spin-exchange rate is consistent with previous measurements \cite{Sikorsky2018}. The prediction of our numerical simulation for $p^\textrm{pass}_\textrm{SE}$ vs. $E_{\textrm{EMM}}$ using this fitted value is shown in Fig.~\ref{fig:SE_EMM} by the solid black line with grey confidence interval around it and is in good agreement with our measured data. The detection efficiency at high EMM is reduced due to coupling of the EMM energy to the secular motion (see \ref{fig:P_shelving_EMM}). The fitted curve includes this effect, calculated numerically, see Methods. The dashed blue line shows the predictions of our model for the endothermic channel given $p_\textrm{SE}^\textrm{0}=0.12$. As seen, our model predicts a much stronger suppression of $p^\textrm{pass}_\textrm{SE}$ at low $E_{\textrm{EMM}}$ than we measure. 

Using the estimated microscopic spin-exchange probability, we can further simulate the dynamics of the spin-exchange processes for different magnetic fields. We present the expected $p^\textrm{pass}_\textrm{SE}$ of the numerical molecular-dynamics simulation in dotted lines in Fig.~\ref{fig:SE_B} with $p^\textrm{0}_\textrm{SE}$ = 0.12, and no fit parameters. The simulation results are in good agreement with the enhancement of exothermic spin-exchange as a function of the magnetic field, and the suppression of the endothermic spin-exchange in the low magnetic field regime. The MD simulations of the endothermic channel underestimate the spin-exchange probability at high magnetic fields compared to the measured data. This difference might result from effects that are not included in the simulation (see Methods).

The formation of trapped-induced bound states is a universal phenomena that can emerge at different trap configurations. For example, in Fig.~\ref{fig:Bound_sim}(c) we show the probability for more than one short-range collision, $p_\textrm{bound}$, as a function of the trap secular frequency $\omega$. Here both for an ideal, time-independent spherical symmetric harmonic trap (blue) and a Paul trap (red) with radial frequency $\omega$ and fixed axial frequency of $3$ kHz. As seen in the figure, the presence of both types of traps leads to the formation of bound states in binary collisions. For the harmonic case the probability to form a bound state increases with the trap frequency to close to unity, but saturates for a Paul trap at $p_\textrm{bound}\simeq 0.5$, owing to work done by the time-dependent fields that heats the ion up and prevents molecular binding \cite{Cetina2012}. The binding probability also depends on the atom-ion mass ratio, as shown in \ref{fig:Bound_vs_m_a}. For the one-dimensional harmonic case, the coupling term or the Hamiltonian is given by $H_\textrm{cpl}=\omega^2 \mu R\cdot r$, where $\mu$ is the reduced mass and $r$ ($R$) is the relative (center-of-mass) coordinate (see Methods for the derivation). This coupling will be negligible for a free ion, or smaller reduced mass, which is qualitatively reproduced in the full simulation results.
These results highlight that the formation of bound state is predominantly associated with the breaking of translation symmetry and not with the presence of time-dependent fields and energy non-conservation of atom-ion collisions in Paul traps.

\section*{Binding Energy and Lifetime of trapped-assisted bound states}    
We can use our measured data to estimate different parameters of the loosely bound molecules. A maximum likelihood estimation of the magnetic dependence of the exothermic channel to an exponential decay gives a decay constant of B$_0$=3.9(7) G, which corresponds to mean bond energy of E$_\textrm{bind}=\Delta_\textrm{SE}B_0=$0.7(1) mK. The likelihood of the exponential decay can be compared to a constant probability model using a likelihood ratio test, giving a p-value of 1.2$\cdot 10^{-5}$. For comparison, in the bound state of neutral atoms due to inelastic confinement-induced resonance, the binding energy is on the order of a few harmonic oscillator quanta \cite{Sala2012}. For our case, $E_\textrm{bind}\cdot k_B/h=15(2)$ MHz, which are $\sim$ 10 quanta of the secular frequency (for an average frequency of 1.2 MHz in the radial modes). However, the RF fields might also play a role in setting the binding energy.

The number of short-range collisions and the resulting molecular lifetime can be roughly estimated from the experiment (i.e.~without a full calculation of the trajectories of the particles) by a simple model for a bound state. Using this model, we find the effect on the probability of spin-exchange in a single pass of the atomic cloud. The bound state can be modeled by assuming that the number of short-range interaction periods, N, has a geometric distribution with a probability 1/$\langle N\rangle$. For any short-range SE probability, $p^\textrm{0}_{\textrm{SE}}$, we can calculate what would be the effective spin-exchange in two regimes: when the process has no energetic gap (zero magnetic field) and when the process is exothermic with an energetic gap larger than the binding energy (large magnetic field), see Methods for details. This amplification depends both on $p^\textrm{0}_{\textrm{SE}}$ and $\langle N\rangle$, and can be extracted from the experiment using an exponential fit to the exothermic channel in Fig. \ref{fig:SE_B}. Combining both measured amplification and the estimation for $p^0_{\textrm{SE}}$ and comparing the model, the mean number of collisions in the experiment is $\langle N \rangle^\textrm{exp} = 8(2)$, see \ref{fig:find_N_exp}. Although the assumptions in this model are rough, this value is similar to the value calculated by the full MD simulation at zero magnetic field, $\langle N\rangle^\textrm{MD}\approx $5.6(2), see \ref{fig:N_mean_vs_B_sim}.

We can see that both results give $p_\textrm{SE}^0\langle N\rangle\approx 1$. If the number of short-range collisions was larger than the exchange probability, i.e. $p_\textrm{SE}^0\langle N\rangle\gg 1$, we would expect the exothermic spin-exchange rate to saturate at the Langevin rate at high magnetic fields and to approach half of the Langevin rate at low fields.

Assuming the binding is only due to the $C_4 / 2r^{4}$ potential, we can estimate the maximal distance between the particles, $r_{max}=\sqrt[4]{C_4/2E_\textrm{bind}}=27$ nm. The falling time of two free particles in $C_4 / 2r^{4}$ potential can be calculated analytically for the one-dimensional problem, and is given by $t\approx1.8\sqrt{\frac{\mu}{9C_4}}r_{max}^3$ (see Methods), which gives a falling time of $t=31$ ns. Therefore, the period of oscillation is $T=62$ ns, and we can estimate the lifetime of the molecule in the experiment by $0.5(1)$ $\mu$s. This estimation is also similar to the mean lifetime of the molecule in the MD simulation, which is $0.411(4)$ $\mu$s.

\section*{Discussion}
We have observed the formation of trap-assisted bound state in ultra-cold atom-ion collisions. These bound states amplify the rate of inelastic processes in both the exothermic and endothermic collision channels. From the measurements of the spin-exchange rates, we estimated the average binding energy of these molecules by 0.7(1) mK and the molecule mean lifetime by 0.5(1) $\mu$s. 

Remarkably, bound states are efficiently formed in binary collisions. Numerical simulations indicate formation every other collision, and molecular life-time with a power-law distribution. Power-laws in the energy distribution were already observed in atom-ion collisions due to multiplicative energy fluctuations after multiple collisions \cite{Meir2016,Rouse2017,Meir2018}. However, in this case, the power-law arises in a single collision. In each Langevin collision, the number of short-range encounters during close contact is deterministic and given by the initial condition.

The molecular binding probability and lifetime can be controlled by the trap parameters, which can serve as a convenient control knob.

We observe that trap-assisted bound states are also formed, and even more efficiently, in static potentials. This indicates that the source for binding is the coupling between relative and center-of-mass motion by the breaking of translational symmetry rather than the presence of time-dependent fields in Paul traps. This effect can also play a role in atom-ion interactions using an optical trap for the ions \cite{Lambrecht2017}.

The formation of bound states introduces a systematic effect that potentially needs to be accounted for in analyzing scattering processes near the s-wave regime \cite{idziaszek2007controlled,Katz2022suppression}. Yet, for a small atom-ion mass ratio, $m_\textrm{a}\ll m_{\textrm{i}}$, and for weak trapping frequencies, the formation of bound states and their systematic effect is suppressed \cite{Feldker2020}.

The formation of bound states by the trap is akin to the mechanism in which ultracold atoms can bind in anharmonic traps, related to inelastic confinement-induced resonances \cite{Sala2012,Sala2013,Lee2022SpinDynamics,Capecchi2022Observation}. Yet, we expect that the bound state formation would qualitatively differ from the neutral atomic case in two main aspects. First, because the external forces trap primarily the ion but not the atom, the translational variance leads to efficient coupling between the center-of-mass motion and the relative motion (see Methods). Therefore, formation of bound states might be pronounced in various hybrid atom-ion experiments. For cold and co-trapped neutral atoms, on the other hand, owing to the nearly-harmonic trapping potential, such bound state formation is suppressed, and strong anharmonicity in the trapping potential is required to break the translational invariance to enable it. Second, the time dependence of the trapping fields in the Paul trap exert work on the ion that shortens the bound state’s lifetime. We therefore expect that ultracold atoms or ions held in energy-conservative optical traps would feature longer bound state lifetimes than the ones we observed.

The formation of bound states can potentially enhance the probability of other inelastic processes or chemical reactions, as was recently suggested by Hirzler et al. \cite{Hirzler2022TrapAssisted}. Yet, owing to the coupling between the center-of-mass and relative motion, energy from the center-of-mass frame can potentially broaden the collision energy distribution and hinder the observation of low-energy resonant processes, such as shape and Feshbach resonances.

Here we have explained our observations using classical simulations and did not include any quantum effects. However the effects of the magnetic field and collision energy on ``sticky collisions'' \cite{Gregory2019} and molecular formation could be related, in a quantum description to wavefunction resonances. Moreover, we see a significant deviation of our measured spin-exchange cross section in the endothermic channel, at low energy and strong magnetic field, from the predictions of our simulations. This could be, for example, due to the effect of multiple overlapping Feshbach resonances at these magnetic fields.

\hfill 

\begin{acknowledgments}
This work was supported by the Israeli Science Foundation and the Goldring Family Foundation. 
\end{acknowledgments}

\subparagraph{Data availability}
Source data are provided with this paper. Other data that support the findings of this study are available from the corresponding author on a reasonable request.

\bibliographystyle{unsrt}
\bibliography{refs}

\clearpage
\setcounter{equation}{0}
\renewcommand{\figurename}{}
\newcounter{extended_data_fig}
\setcounter{extended_data_fig}{1}
\renewcommand{\thefigure}{Extended Data Fig.~\arabic{extended_data_fig}}
\setcounter{table}{0}

\part*{\centerline{Methods}}
\section*{Experimental sequence}
A cloud of $^{87}\textrm{Rb}$ atoms is trapped and cooled down using a magneto-optical trap (MOT), followed by a dark-MOT stage and a polarization gradient cooling. A cloud of $\sim 10^6$ atoms is loaded into an optical lattice created by counter-propagating $1064$ nm laser beams. The atoms are pumped into one of the hyperfine states $|1,-1\rangle$ or $|1,+1\rangle$ by a sequence of MW and optical pulses. After pumping, more than 90\% of the population is in the desired spin state.

The ion is trapped in a linear segmented Paul trap where the secular frequencies are $f=(1.1,1.3,0.48)$ MHz, and the RF frequency is $\Omega_\textrm{RF}=2\pi\cdot 26.51$ MHz. The ion is Doppler cooled and subsequently cooled near its ground state with $\bar{n}<0.5$ in each of the motional modes. The atoms are moved to the lower chamber by changing the relative frequency between lattice beams. Unless stated otherwise, the atoms are moved over the ion in a velocity of 0.14 m/s, which is equivalent to a kinetic energy of 100 $\mu K$. The current in the quantization coils is ramped up or down after cooling of the ion and 30 ms before the collision. Due to eddy currents in the system, the magnetic field has a transient time of 5 ms. The current in the quantization coils is returned to its initial value after the collision.

The detection of the spin state was done 70 ms after the collision ends and consists of two stages. First, we apply two $\pi$-pulses from the spin state that was not the initial state to two different states in the D$_{5/2}$ manifold (double shelving). For example, in the exothermic transition, the ion is initialized at $|+1/2\rangle$, the detection pi pulses are $|-1/2\rangle \to |-5/2\rangle$ and $|-1/2\rangle \to |+3/2\rangle$. After the shelving pulses, a fluorescence detection scheme is applied to check whether the ion spin was changed or not. If spin exchange did not occur, the ion remains in the S$_{1/2}$ manifold and will appear bright. Otherwise, it is shelved into the D$_{5/2}$ manifold and no photons are detected. The same detection scheme is applied for the same experimental sequence, but without atoms, to detect the preparation efficiency of the ion's spin state, which is independent of the magnetic field and EMM, and is about 0.5\%. Experiments are taken interlacing over the scanned variables and the background experiments.

\section*{Measuring the Langevin collisions probability}
For two free particles, the Langevin rate is given by
\begin{equation}
\Gamma_L=n\sigma_L(E_\textrm{col})\sqrt{\frac{2E_\textrm{col}}{\mu}},
\end{equation}
where $n$ is the atomic density, $\sigma_L$ is the total Langevin cross-section, $E_\textrm{col}$ is the energy of the collision in the center-of-mass frame, and $\mu$ is the reduced mass. Classically, the cross-section is given by $\sigma_L=\pi\sqrt{2C_4/E_\textrm{col}}$, and therefore the reaction rate coefficient is
\begin{equation}
    K_\textrm{L}=\sigma_L(E_\textrm{col})\sqrt{\frac{2E_\textrm{col}}{\mu}}=\pi \sqrt{\frac{4C_4}{\mu}},
\end{equation}
which is independent of the collision energy.

In the experiment, the atomic density in the ion position, $n$, is time-dependant due to the movement of the lattice \cite{lattice2021}, and the number of Langevin collisions per passage is given by
\begin{equation}
    N_\textrm{L}=K_\textrm{L}\int_{-\infty}^\infty n(t)dt.
\end{equation}

Assuming the lattice is moving in a constant velocity, $v_\textrm{lattice}$, the number of collisions can be written as a function of the atomic density integrated along the direction of motion, $\rho$,
\begin{equation}
    N_\textrm{L}=\frac{K_\textrm{L}}{v_\textrm{lattice}}\int_{-\infty}^\infty n(x)dx\equiv \frac{\rho K_\textrm{L}}{v_\textrm{lattice}},
    \label{Eq:N_col_pass}
\end{equation}
where $x$ is the position of the ion relative to the center of the atomic cloud.

We measured the number of collisions per passage by measuring the failure probability to shelve the ion from the electronic ground state to the D level using a long shelving pulse ($14.6\,\mu$s) in the presence of high EMM energy ($E_{\textrm{EMM}}=2.5$ K) as described in Ref.~\cite{Katz2022suppression}. Under these conditions, a Langevin collision will efficiently couple the EMM energy to the secular motion \cite{Zipkes2011,chen2014neutral}, and by that decrease the shelving probability. We numerically simulate this process \cite{Katz2022suppression}, and find that the expected probability for a shelving failure per Langevin collision is about $87\%$.

\begin{figure}[b]
    \centering
    \includegraphics[width=8.6cm]{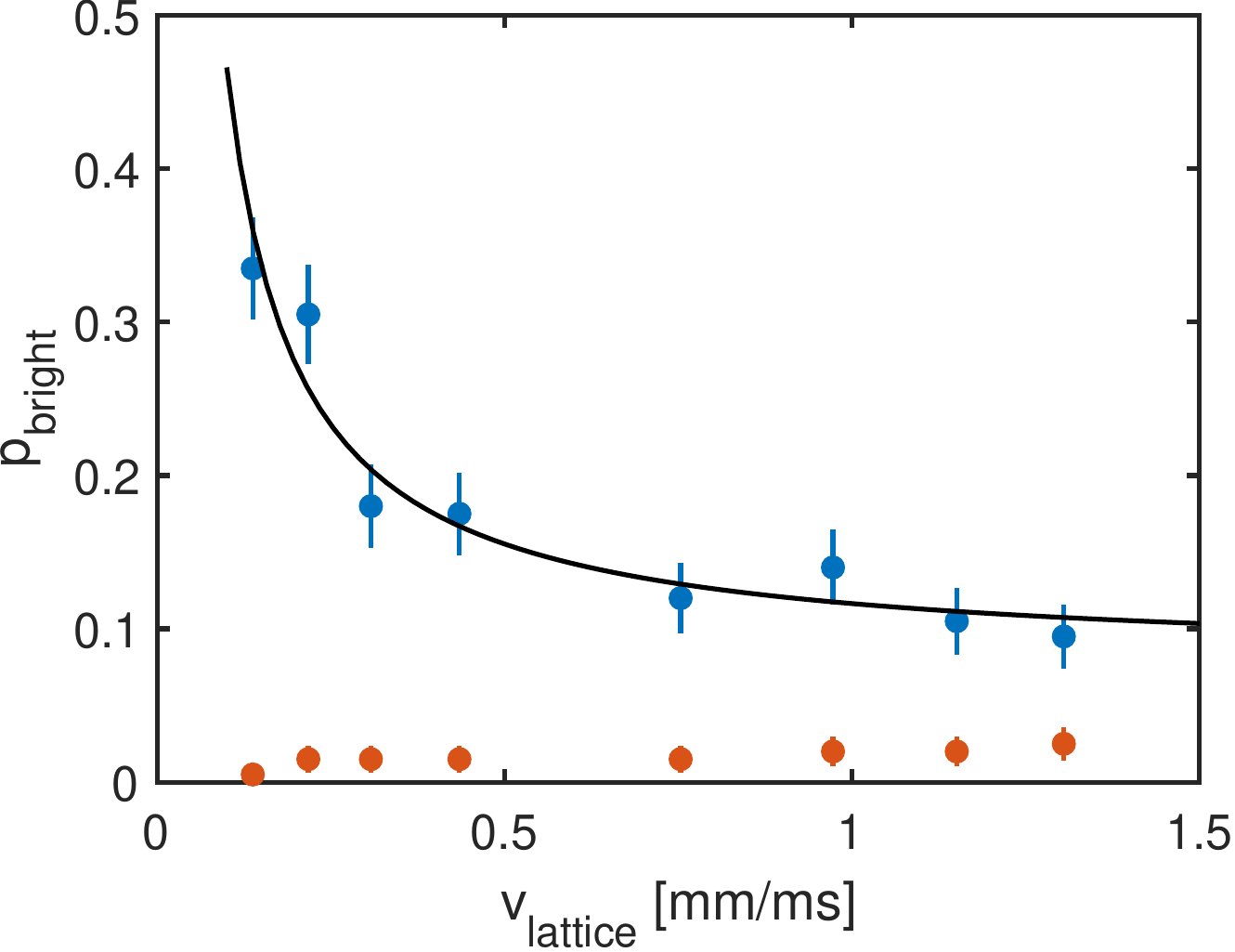}
    \caption{\textbf{Calibration of the number of Langevin collisions.} The probability of observing the ion in a bright state after double shelving pulses with atoms (blue) and without (red). When atoms are present, this probability is proportional to the probability of at least one Langevin collision in a lattice passage. Solid line is a fit to Eq.~\ref{Eq:N_col_pass}, with $\rho K_\textrm{L}$ = 0.039(3) and $p_\textrm{bg}$ = 0.078(8). Error bars are binomial distribution standard deviation.} 
    \label{fig:Langevin_calib}
\end{figure}

We measure the shelving-failure probability per passage of the cloud and present it in \ref{fig:Langevin_calib} as the probability of observing the ion in a bright state after two shelving pulses as a function of the velocity of the ultracold atomic cloud. We find that the probability fits well to
\begin{equation}
p_\textrm{bright}=\rho K_\textrm{L}/{v_\textrm{lattice}} +p_{\textrm{bg}},    
\end{equation}
where the first term is expected from Eq.~\ref{Eq:N_col_pass}, and the second term describes a constant background. From maximum likelihood estimation, $\rho K_\textrm{L}=0.039(3)$ and $p_\textrm{bg}=0.078(8)$. The background probability is not due to finite shelving efficiency (which is $98.4(3)\%$), but likely due to hot atoms that are not trapped in the lattice. Taking into account the detection efficiency of a collision, at a collision energy of 100 $\mu K$, the probability for a cold Langevin collision in a single pass is $p_\textrm{L}^\textrm{lattice}$=0.32(3) with a background of about $p_\textrm{L}^\textrm{bg}=$0.089(4). This value corresponds to the probability to have at least one collision in a pass. Assuming the number of collision in a single pass has a Poisson distribution, the mean number of collisions per pass is $\langle N\rangle$=0.385, with a probability of $p_\textrm{L}^\textrm{single}$=0.26 for a single collision and $p_\textrm{L}^\textrm{multiple}$=0.06 for multiple collisions.

\stepcounter{extended_data_fig}

\section*{Rabi Carrier Thermometry}
The temperature of the ion after a collision is measured by Rabi carrier thermometry \cite{Meir2016,Pinkas2020}. The probability of observing the ion in the $D_{5/2}$ manifold after a resonant pulse driving the $S_{1/2}\to D_{5/2}$ transition with duration $t$ is given by \cite{Leibfried2003}
\begin{equation}
    P_D(t)=\sum_\bold{n}P(\bold{n})\sin^2(\Omega_{\bold{n},\bold{n}}),
\end{equation}
where $P(\bold{n})$ is the occupation probability of the Fock state $\bold{n}$. Here $\Omega_{\bold{n},\bold{n}}$ the coupling strength between $|S_{1/2}\rangle|\bold{n}\rangle$ and $|D_{5/2}\rangle|\bold{n}\rangle$ is given by \cite{Leibfried2003}
\begin{equation}
    \Omega_{\bold{n},\bold{n}}=\Omega_0 \prod_i e^{\eta_i^2/2}L_{n_i}(\eta_i^2 ),
\end{equation}
where $\Omega_0$ in the scaled interaction strength, $\eta_i$ is the Lamb-Dicke parameter of the $i$-th mode, and $L_n(x)$ is the n-th Laguerre polynomial. 
The Fock states are assumed to be distributed thermally with a mean occupation number $\bold{\bar{n}}$, 
\begin{equation}
    P(\bold{n};\bold{\bar{n}})=\prod _i \frac{1}{\bar{n}_i+1} \left(\frac{\bar{n}_i}{\bar{n}_i+1}\right)^{n_i}.
\end{equation}

At higher temperatures, more values of n need to be considered in this distribution. At temperatures above 2 mK, we approximate the energy distribution by
\begin{equation}
    P(E)=\frac{1}{(k_BT)^3}E^2 e^{-\frac{E}{k_BT}}
\end{equation}
where $E=\sum_i \hbar \omega_i n_i$ and $n_i$ is taken over a logarithmic scale.
The contrast of a Rabi cycle post-selecting the spin-state proportional to the spin-exchange probability is given by
\begin{equation}
    P_D(t;p_{SE})=p^\textrm{pass}_{\textrm{SE}}P_D(t).
\end{equation}
The parameters $T$, $\Omega_0$ and $p^\textrm{pass}_{\textrm{SE}}$ are found by maximum-likelihood estimation of the experimental results to $P_D(t;p_{SE})$ assuming that the number of dark events follows a binomial distribution.

\begin{figure}
    \centering
    \includegraphics{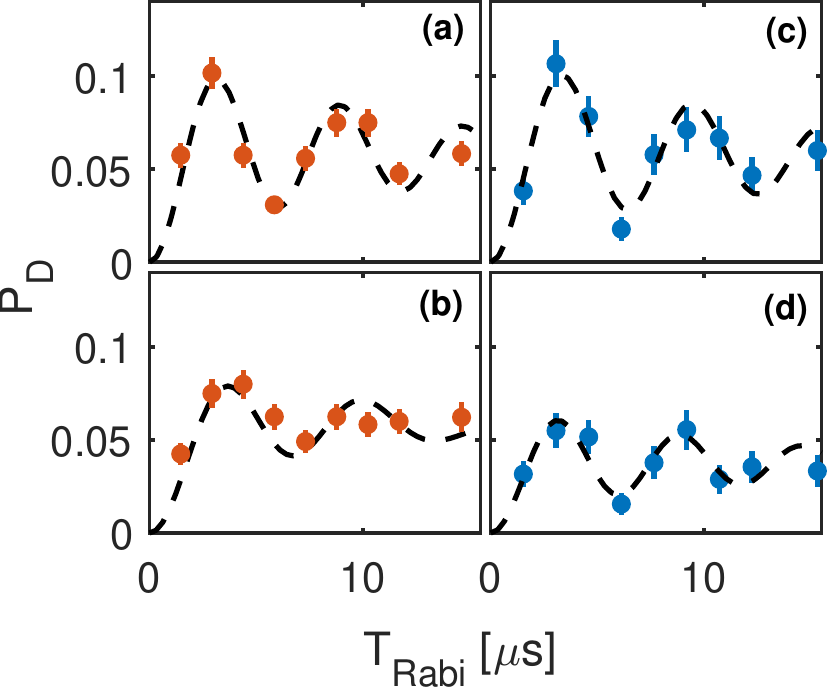}
    \caption{\textbf{Rabi carrier thermometry after post-selecting SE events.} (a-b) exothermic transitions at 3 G (a) and 20 G (b). (c-d) endothermic transitions at 3 G (c) and 20 G (d). Temperatures and contrast of the Rabi oscillation are written in Table \ref{tab:T_p_se}. Error bars are 1$\sigma$ binomial standard deviation.}
    \label{fig:my_label}
\end{figure}
\stepcounter{extended_data_fig}

\begin{table*}
\begin{tabular}{|c | c || c | c | c | c |} 
 \hline
 Channel & B [G] & T (atoms) [mK] & Contrast (atoms) & T (bg) [mK] & Contrast (bg) \\ [0.5ex] 
 \hline\hline
 Exothermic & 3 & 1.3(5) & 11.4(4) & 0.24(3) & 0.98(2) \\ 
 \hline
 Exothermic & 20 & 4.4(4) & 12.3(5) & 0.41(5) & 0.97(2) \\
 \hline
 Endothermic & 3 & 1.2(2) & 11.4(6) & 0.21(2) & 0.99(2)\\
 \hline
 Endothermic & 20 & 1.5(4) & 7.4(5) & 0.24(4) & 0.96(2) \\
 \hline
\end{tabular}
\caption{Temperatures and contrasts fitted to Rabi cycles for different spin-exchange channels with the corresponding temperature and contrast measured on the same transition without atoms.}
\label{tab:T_p_se}
\end{table*}

\section*{Avoided-crossing due to trapping RF field}
Due to the oscillating electric field, there is an oscillating magnetic field on the ion with the same frequency, 26.5 MHz. This creates an avoided crossing when the Zeeman gap in the ground state of the ion is equal to this frequency, at 9.5 G. As a result, if the ion is prepared in $|\downarrow\rangle$ and the magnetic field during the collision is above this threshold, the collision will occur when the ion is in the $|\uparrow\rangle$, and similarly for the opposite state. Therefore, for measurements above this threshold with ion in $|\uparrow\rangle$, the ion should be prepared in the $|\downarrow\rangle$ state, and vice versa.

We can estimate the fraction that does not follow the adiabatic transition, i.e. remains in the initial state, by the Landau-Zener formula \cite{SteckQuantumAtom},
\begin{equation}
    P_{lost}=\exp\left(-\frac{\pi \Omega^2}{2|\partial_t\Delta|}\right),
\end{equation}
where $\Omega$ is the Rabi frequency of the transition and $\partial_t\Delta$ is the rate of the frequency sweep. The magnetic field is ramped up in $5$ ms from $3$ G to a maximal value of $20$ G. This gives a maximal frequency velocity of $\partial_t\Delta=9.5\cdot 10^{-3} \,\text{MHz}^2$. The amplitude of the oscillating magnetic fields can be estimated from the frequency dependence of the EMM compensation voltage for different Zeeman transitions \cite{Meir2018Apparatus}. For the $\left|-1/2\right\rangle \to \left|-5/2\right\rangle$ transition, the oscillating magnetic field is equal to the EMM amplitude at $u_{\textrm{EMM}}\approx$ 1 nm, which corresponds to a Doppler shift of about $\Delta f =k\cdot u_{\textrm{EMM}}\Omega_{\textrm{RF}}=1.55$ MHz, where $k$ is the wave-number of the 674 nm laser. The magnetic field that creates this modulation is given by $B=\frac{\Delta f}{2.8 \textrm{MHz}}=0.55$ G, which gives a Rabi frequency of $\Omega^2=2.37\ (\textrm{MHz})^2$. Since $\Omega^2\gg |\partial_t\Delta|$ the adiabatic condition is satisfied and the remaining population should be negligible. No significant population is measured in background experiments without atoms (about 0.5 \%).

\section*{EMM energy calibration}
We control the EMM energy by changing the voltage difference $\Delta V_{\textrm{comp}}$ on an additional electrode, relative to the compensated value, generating a static electric field that shifts the equilibrium position of the ion away from the RF-null of the Paul trap's potential. Consequently, the ion is subject to the trap's electric field in the radial direction that oscillated at the RF frequency $\Omega_{\text{RF}}$, generating EMM displacements at the same frequency with amplitude $\bold{u}_\textrm{EMM}$. The EMM energy the ion gains is therefore \begin{equation}\label{eq:EMM_energy}E_\textrm{EMM}=\frac{m_i(u_\textrm{EMM}\Omega_{\text{RF}})^2}{4}.\end{equation} 

We measure the EMM oscillation amplitude as a function of $\Delta V_\textrm{comp}$ by following the protocol in Refs.~\cite{Meir2018Apparatus,Meir2016DynamicsSingle}, comparing $\Omega_0$, the Rabi frequency the carrier $S_{1/2}-D_{5/2}$ transition, with $\Omega_\textrm{sb}$, the Rabi frequency of the first micromotion sideband transition (i.e. a transition detuned from the carrier transition by the micromotion frequency $\Omega_{\text{RF}}$). The ratio of these two Rabi frequencies is related to $u_\textrm{EMM}$ by 
\begin{equation}\label{eq:Ziv_micromotion}
\frac{\Omega_\textrm{sb}}{\Omega_0}=\frac{\bold{k}\cdot \bold{u}_\textrm{EMM}}{2},
\end{equation}
where $\bold{k}$ is the wave-vector of the narrow linewidth laser at 674 nm that we used to drive the transition. Eq.~(\ref{eq:Ziv_micromotion}) holds for weak micromotion amplitudes $\bold{k}\cdot \bold{u}_\textrm{EMM}\ll1$ \cite{Meir2018Apparatus}, which is valid in our experiment.  

\begin{figure}
    \centering
    \includegraphics[width=8.6cm]{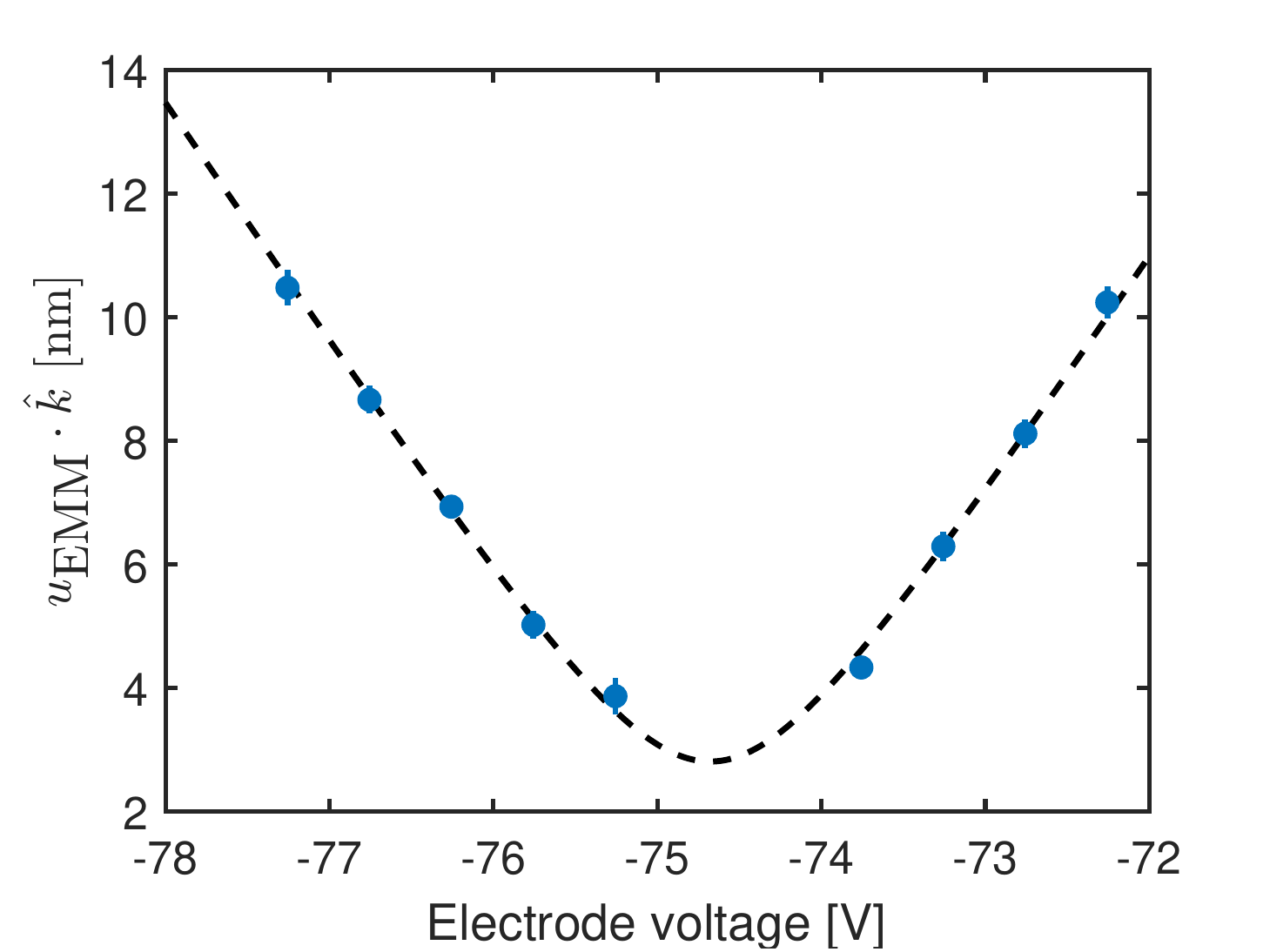}
    \caption{\textbf{Calibration of the projection of the EMM amplitude along the shelving beam axis as a function of the applied voltage on an external electrode.} Measured data (blue circles) is extracted from Eq.~\ref{eq:Ziv_micromotion} and dashed line are a fit to Eq.~\ref{eq:EMM_calib_fit}, where $\gamma_{\textrm{proj}}=3.97(6)$ nm/V, $V_0=-74.68(2)$ V, and $c=2.8(2)$ nm with 1$\sigma$ confidence bounds.}
    \label{fig:EMM_V_calib}
\end{figure}

We measure $\Omega_\textrm{sb}$ and $\Omega_\textrm{0}$ as a function of $\Delta V_{\textrm{comp}}$ and show the projection of the micromotion amplitude along the optical axis $\hat{k}$ in \ref{fig:EMM_V_calib} using Eq.~(\ref{eq:Ziv_micromotion}). We fit the data to the following function
\begin{equation}
    \bold{u}_\textrm{EMM}\cdot\hat{\bold{k}}=\sqrt{\gamma_\textrm{proj}^2(V-V_0)^2+c^2}
    \label{eq:EMM_calib_fit}
\end{equation}
where $V$ is the applied voltage, $V_0$ is the value at which micro-motion is compensated and $\Delta V_{\textrm{comp}}=V-V_0$. The coefficient $\gamma_\textrm{proj}$, the offset $c$, and $V_0$ are fitting parameters of this model. We use the parameter $c$ because the estimation of $\Omega_\textrm{sb}$ is limited by the coherence time of the system \cite{Meir2016DynamicsSingle}. However, the response of the electrode is given solely by $\gamma_\textrm{proj}$. We find $\gamma_{\textrm{proj}}=3.97(6)$ nm/V, $V_0=-74.68(2)$ V, and $c=2.8(2)$ nm with 1$\sigma$ confidence bounds. We extract the total micromotion amplitude by accounting for the relative orientation of $\hat{k}$ with respect to the trap axes. In our trap apparatus $\hat{k}=\frac{1}{\sqrt{2}}(\hat{y}+\hat{z})$ and $\hat{u}_\textrm{EMM}=\cos(19^\circ)\hat{y}+\sin(19^\circ)\hat{x}$ 
 as was previously measured in Ref.~\cite{Meir2016DynamicsSingle}, using the trap coordinates $\hat{z}$ for the axial and $\hat{x},\, \hat{y}$ for the radial directions. The geometric factor is therefore $\hat{k}\cdot\hat{u}_\textrm{EMM}=0.667$. We can then calibrate the dependence of the excess micromotion energy in Eq.~\ref{eq:EMM_energy} on the applied voltage in Eq.~\ref{eq:EMM_calib_fit}, yielding $E_{\text{EMM}}/\Delta V_{\textrm{comp}}^2=2.58$ mK/V$^2$.

\stepcounter{extended_data_fig}
\section*{Molecular dynamics simulation}
Based on Refs.~\cite{Cetina2012,Meir2018,Pinkas2020}, we solve Newton's equations of motion for the atom and ion Hamiltonian,

\begin{equation}
\begin{split}
    H &=\frac{\hat{\bold{p}}_\textrm{i}^2}{2m_\textrm{i}}+\frac{m_\textrm{i}\Omega^2_{\textrm{RF}}}{8}\sum_j\left[a_j+2q_j\cos(\Omega_{\textrm{RF}}t) \right]\hat{r}_{\textrm{i},j}^2\\
      &+ \frac{\hat{\bold{p}}_\textrm{a}^2}{2m_\textrm{a}}+V(|\hat{\bold{r}}_\textrm{i}-\hat{\bold{r}}_\textrm{a}|)
\label{eq:Hamiltonian}      
\end{split}
\end{equation}

where $m_{\textrm{i}(\textrm{a})}$ is the ion (atom) mass, $r_{\textrm{i(a)},j}$ is the ion (atom) position in the $j$-th direction, $V(|\hat{\bold{r}}_\textrm{i}-\hat{\bold{r}}_\textrm{a}|)$ is the atom-ion interaction potential, $\Omega_{\textrm{RF}}$ is the RF frequency, and $a_j$ and $q_j$ are the trap parameters in the $j$-th direction. Unless stated otherwise, $a=(-2.2,0.83,1.3)\cdot 10^{-3}$ and $q=(-0.134,0.134,0)$ reproducing the secular trapping frequencies of $f=(1.1,1.3,0.48)$ MHz, as in the experiment. 
We use the asymptotic form of the atom-ion interaction potential $V(r)=-{2C_4}/{r^4}$ at all distances larger than $10$ nm, where the coefficient $C_4=1.09\cdot10^{-56}$ J$\cdot$m$^4$ is determined predominantly by the neutral atom polarizability. The short-range atom-ion potential is modeled by an infinite wall at 10 nm. At this point, the particles collide in an elastic hard-sphere collision. Without spin-exchange, after the collisions the velocities in the collision axis are changed assuming conservation of momentum and energy in the center-of-mass frame. Because the masses of $^{87}$Rb and $^{88}\textrm{Sr}^+$ are nearly equal, the velocities are as well nearly equal but change their sign in the center-of-mass frame.

The atoms start their trajectory on a plane 1.2 $\mu$m from the center of the trap and the ion has an initial temperature of 150 $\mu$K (unless stated otherwise) divided equally in the different secular motional modes. The velocity distribution of the atoms consists of two terms, a thermal term (T = 5 $\mu K$) and a constant velocity which corresponds to the velocity of the optical lattice. For each set of parameters, the total number of calculated trajectories is $10^4$, out of which about 40 \% have at least one short-range collision. The confidence bounds of the bound state and spin-exchange probabilities are $1\sigma$ of a binomial distribution.

The kinetic effect of a spin exchange process is modeled by energy absorption or release after the hard-sphere collision. The probability of a spin-exchange to happen is calculated by acceptance-rejection method, for the given $p^\textrm{0}_\textrm{SE}$ value. If spin-exchange occurred, the momentum and energy is no longer conserved. The momentum is altered only along the collision axis, and the ion's and atom's velocities parallel to the collision axis after the collision are updated by
\begin{equation}
\begin{split}
u_{i}^{\parallel}=v_{cm}^{\parallel}+\frac{1}{m_{i}}\sqrt{\frac{2\Delta E}{\mu}+\left(\frac{|v_{i}-v_{a}|}{\mu}\right)^{2}}, \\
u_{a}^{\parallel}=v_{cm}^{\parallel}-\frac{1}{m_{a}}\sqrt{\frac{2\Delta E}{\mu}+\left(\frac{|v_{i}-v_{a}|}{\mu}\right)^{2}}. \\
\end{split}
\end{equation} where $\Delta E=h B\cdot 3.5$ MHz, $h$ is Planck constant, $v_{cm}^{\parallel}$ in the center-of-mass velocity parallel to the collision axis, and 3.5 MHz/G is the Zeeman energy gap in a spin-exchange transition when the ion and the atom are in the electronic ground state.

\subsection*{Calculating effective spin-exchange in the presence of EMM}
An excess micromotion induced by a constant field can be introduced into the simulation by an additional acceleration term,

\begin{equation}
\bold{\ddot{x}}=\frac{e\bold{E}^{\textrm{dc}}}{m_i}
\end{equation}
where $e$ is the electron charge. Only radial EMM is assumed, divided equally between the radial modes. For EMM temperature $T_\textrm{EMM}$ in the j-th mode, the electric field is, 
\begin{equation}
    E^\textrm{dc}_j=2\sqrt{\frac{4T_\textrm{EMM}}{m_i\Omega^2}}\frac{m_i\omega_j^2}{q_je}.
\end{equation}

We then calculate numerically the probability density function of the number of short-range collisions, $n$, for a given $E_\textrm{EMM}$, $p(n,E_\textrm{EMM})$. This probability is calculated from 10$^4$ trajectories for each EMM energy, $E_\textrm{EMM}$.
In the case of exothermic transition in the presence of large enough magnetic field, we can assume a spin-exchange event happens at most one time. The effective SE probability per Langevin collision is given by

\begin{equation}
p_{\textrm{SE}}^{\textrm{eff}}(E_\textrm{EMM};p^\textrm{0}_\textrm{SE})=\sum_{n=1}^\infty p(n,E_\textrm{EMM})(1-p^0_{\textrm{SE}})^{n-1}p^0_{\textrm{SE}}.
\label{eq:p_SE_eff_MD_sim}
\end{equation}

\begin{figure}
    \centering
    \includegraphics[width=1\linewidth]{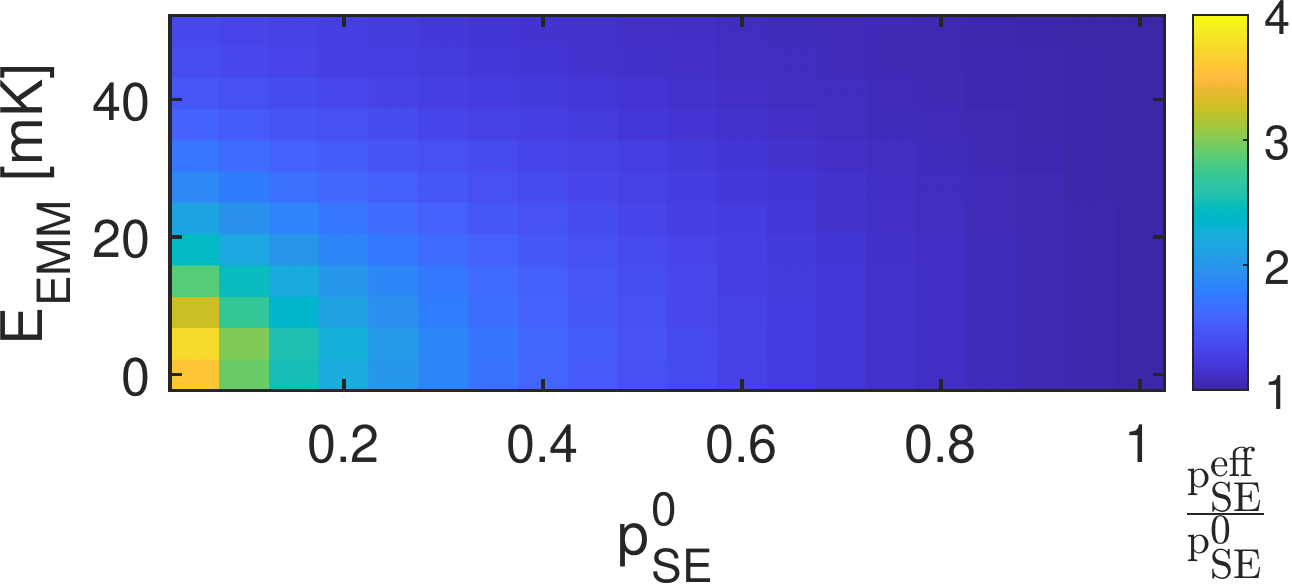}
    \caption{\textbf{Amplification of the short-range spin exchange, based on the MD simulation.} The amplification, $p_\textrm{SE}^\textrm{eff}/p_\textrm{SE}^\textrm{0}$, is calculated by Eq.~\ref{eq:p_SE_eff_MD_sim} for different EMM energies, $E_\textrm{EMM}$, and short-range spin-exchange probabilities, $p_\textrm{SE}^\textrm{0}$, at zero magnetic field.
    }
    \label{fig:p_SE_EMM_amplification}
\end{figure}

From this result, we calculate the amplification factor, $p_\textrm{SE}^\textrm{eff}/p_\textrm{SE}^\textrm{0}$, shown in \ref{fig:p_SE_EMM_amplification} for different values of EMM energy and short-range spin exchange.

Since the probability for a collision in a single passage, $p_\textrm{L}^\textrm{lattice}$, is less than one, and there are contributions from hot collisions with probability $p_\textrm{L}^\textrm{bg}$, the effective observed probability from the simulation is given by
\begin{equation}
    p_\textrm{SE}^\textrm{pass}(E_\textrm{EMM};p^\textrm{0}_\textrm{SE})=p_\textrm{L}^\textrm{lattice}p_\textrm{SE}^\textrm{eff}+p_\textrm{L}^\textrm{bg}p_\textrm{SE}^0.
\end{equation}
Using maximum likelihood estimation, we find the $p^0_{\textrm{SE}}$ that maximizes the likelihood function.

\stepcounter{extended_data_fig}

Although the EMM is compensated throughout the experiment, the measured data is not centered around the zero voltage on the compensation electrode. Therefore, the center of the curve was found before finding the short-range $p^\textrm{0}_\textrm{SE}$. The center was found by a parabola fit to $p^\textrm{pass}_\textrm{SE}$ as function of $\Delta V_\textrm{comp}$. The center is shifted by 0.2073 V on the compensation electrode, corresponding to 50 $\mu K$.

\begin{figure}
    \centering
    \includegraphics[width=1\linewidth]{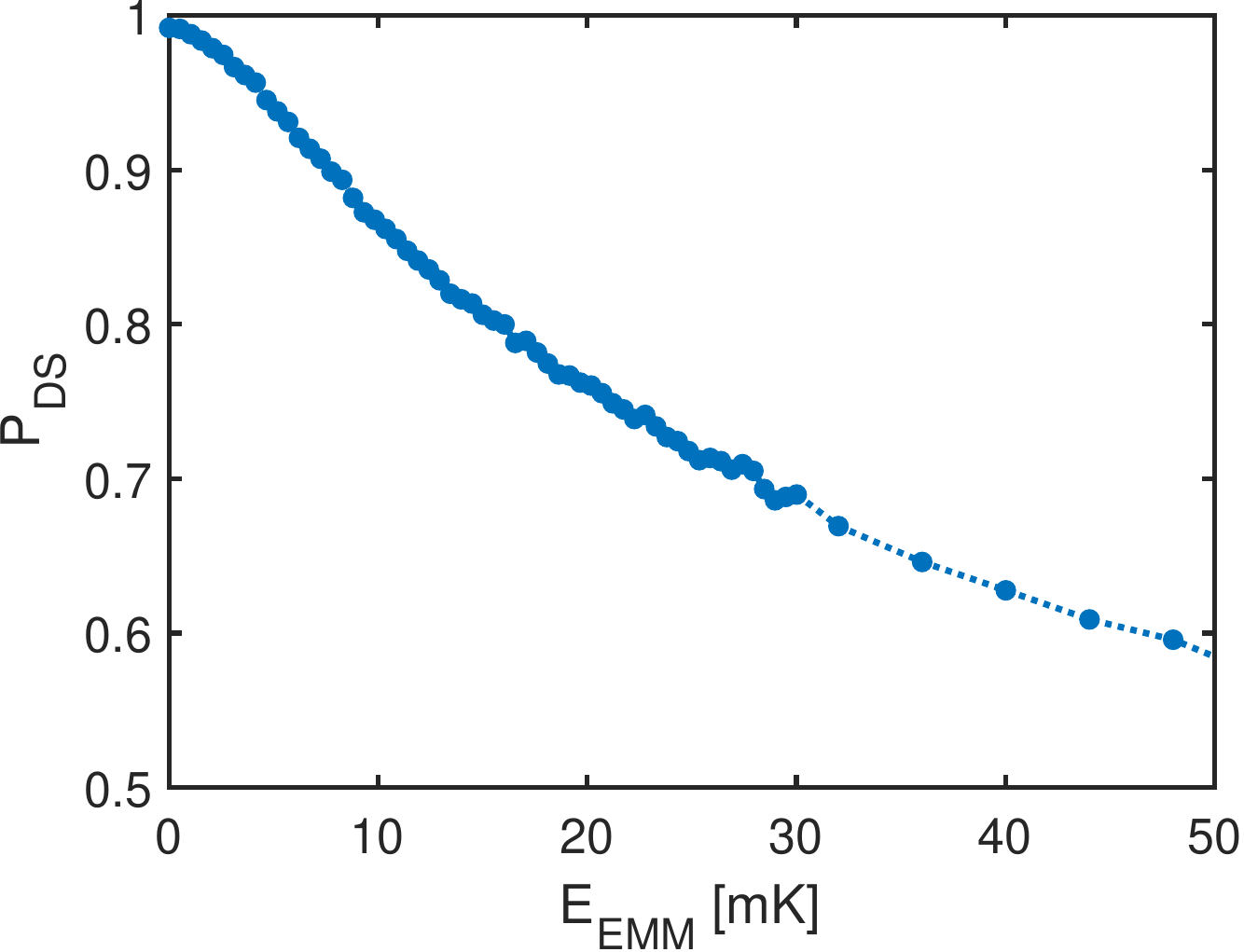}
    \caption{\textbf{Double shelving (DS) efficiency after a collision in presence of EMM.} For each EMM energy, the shelving  probability is calculated by averaging $10^4$ single collision events. Exothermic reaction releasing 2.7 mK (corresponding to the energy gap at 16 G), happens after each collision.}
    \label{fig:P_shelving_EMM}
\end{figure}

Since the EMM can be coupled to the secular motion in a collision, it can heat up the ion. This effect is modeled by finding the energy distribution of the ion after a collision in the presence of specific EMM and then calculating the shelving probability $P_D$ as in Ref.~\cite{Katz2022suppression}, with the parameters of the electroode that was used in this experiment. In the experiment, two shelving pulses are applied, and therefore the double shelving probability is $P_{DS}=P_D(2-P_D)$. $P_{DS}$ as a function of EMM energy for an exothermic transition at 16 G is shown in \ref{fig:P_shelving_EMM}. This spin-exchange probability from the simulation results is multiplied by the corresponding shelving efficiency and presented in Fig.~\ref{fig:SE_EMM} in the main text.

\stepcounter{extended_data_fig}

\begin{figure}
    \centering
    \includegraphics[width=1\linewidth]{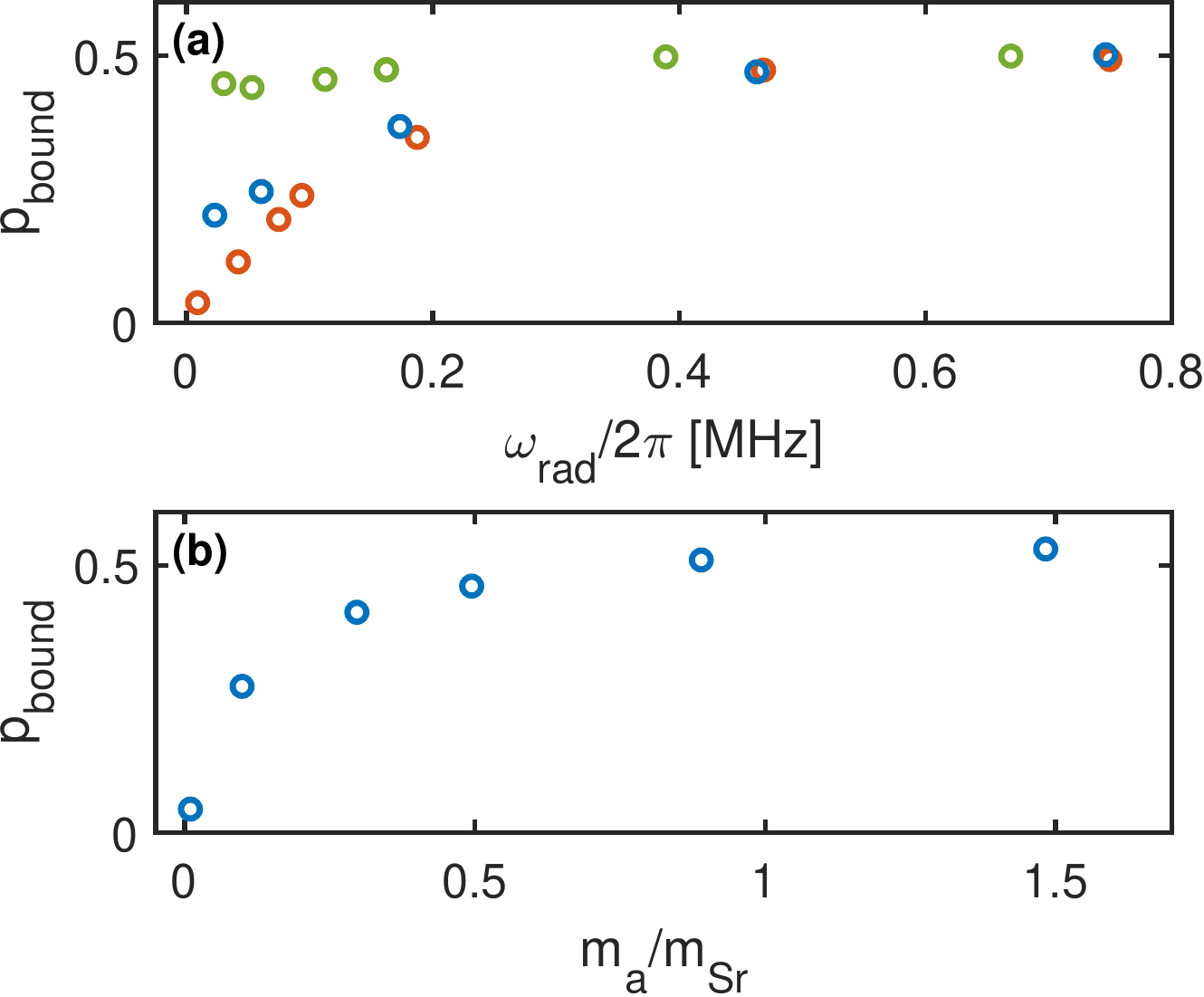}
    \caption{\textbf{Bound state probability in a Paul trap for different axial frequencies and atom's mass.} (a) Bound state probability as a function of radial trap frequency for axial frequency of $\omega_\textrm{ax}/2\pi=3,\ 100,\ 480 $ kHz (red, blue, and green, respectively). The 3 kHz graph is the same as in Fig.~\ref{fig:Bound_sim}. (b) Bound state probability as a function of the atom's mass. All parameters, apart from the atom's mass, are taken as in the experiment. The mass of the atom is changed without changing the polarization constant, $C_4$. For both graphs, each point corresponds to $10^4$ trajectories. Confidence bounds of 1$\sigma$ are on the order of marker size.}
    \label{fig:Bound_vs_m_a}
\end{figure}
\stepcounter{extended_data_fig}

\section*{Deviation of the model from the measured endothermic spin-exchange in high magnetic field}

The MD simulation results deviated from the measured spin-exchange in the endothermic transition at high magnetic field and low collision energy (Figs.~\ref{fig:SE_EMM} and \ref{fig:SE_B} in the main text). This difference might result from weaker effects that are not included in the simulation. 

\begin{figure}
    \centering
    \includegraphics[width=8.6cm]{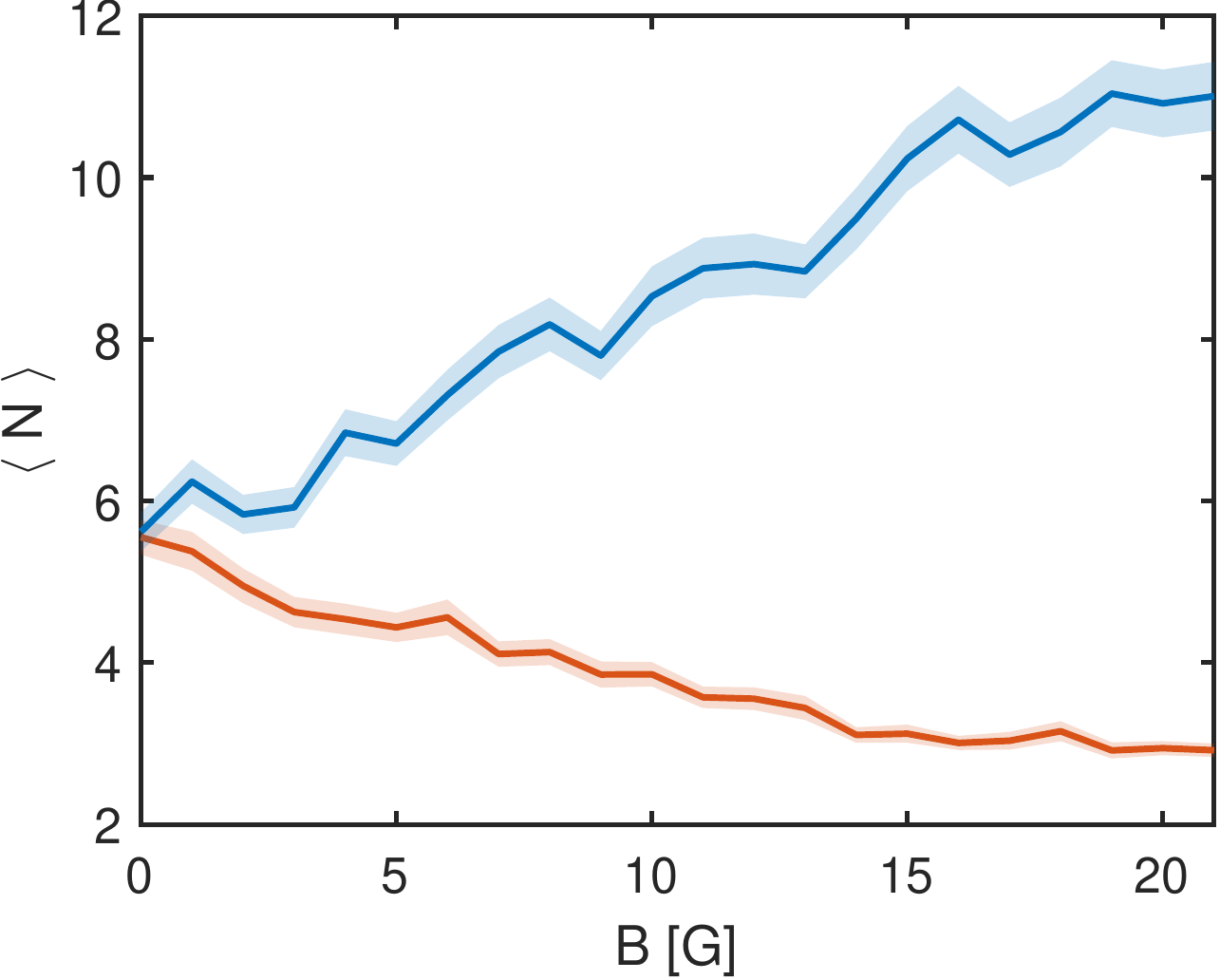}
    \caption{\textbf{Mean number of short-range collisions in a bound state as a function of the magnetic field as calculated by the MD simulation.} The mean number of collisions is calculated for the endothermic (blue), and exothermic (red) transitions, given short-range spin-exchange probability of $p_\textrm{SE}^\textrm{0}=0.12$. Error bars are one standard deviation calculated by bootstrapping the data-set 10 times its size.}
    \label{fig:N_mean_vs_B_sim}
\end{figure}

The MD simulation assumes that spin exchange is the only process in a collision. However, the ion spin can flip due to spin-orbit coupling (spin-relaxation). This process is five times smaller than the exchange in our system \cite{Sikorsky2018} and is not included in the MD simulation. This process has a smaller energetic barrier ($\Delta_{SR}=h\cdot2.8 \frac{MHz}{G}= 134 \frac{\mu K\cdot k_B}{G}$), and therefore should also be suppressed in the high magnetic field. Imperfections in the atoms' spin preparation could lead to the spin change of the ion only by the spin-relaxation channel, and hence should be energetically suppressed. However, the MD simulation predicts a larger number of short-range collisions in the endothermic channel at the large magnetic field due to deeper binding energy after a spin-exchange, as can be seen in \ref{fig:N_mean_vs_B_sim}. This can amplify weaker channels that were neglected in the simulation.

Due to a small but nonzero probability for secondary Langevin collisions, the ion can be heated, which could restrain the suppression of the magnetic barrier. However, previous measurements in our system \cite{Meir2016,Pinkas2020} show that the heating in a single collision is around 100’s $\mu$K, which is still much lower than the energy gap.

In addition, the model assumes that there is a scale separation between the mechanics (the bound state) and the chemistry (spin-exchange), and that the spin-exchange probability is independent of the magnetic field. These assumptions are approximations and might break when there are many short-range collisions. In addition, the cross-section could depend on the magnetic field by Feshbach resonances \cite{tomza2019cold}.

\stepcounter{extended_data_fig}

\section*{A simplified model for the bound state}
In this section, we describe a simple model for the bound state that describes the amplification of the observed exothermic spin-exchange probability in two regimes, zero and infinite magnetic fields. We use this model to get an estimation for the mean number of short-range collisions and the lifetime of the bound state directly from the experiment, without a full calculation of the trajectories of the particles, in contrast to the MD simulation.

Assume that a bound state consists of $n$ short-range collisions, where $n$ has a geometric distribution with a probability $p=1/ \langle N \rangle$. For each short-range collision, the SE probability, $p^\textrm{0}_{\textrm{SE}}$, is constant.

\begin{figure}[t]
    \centering
    \includegraphics[width=1\linewidth]{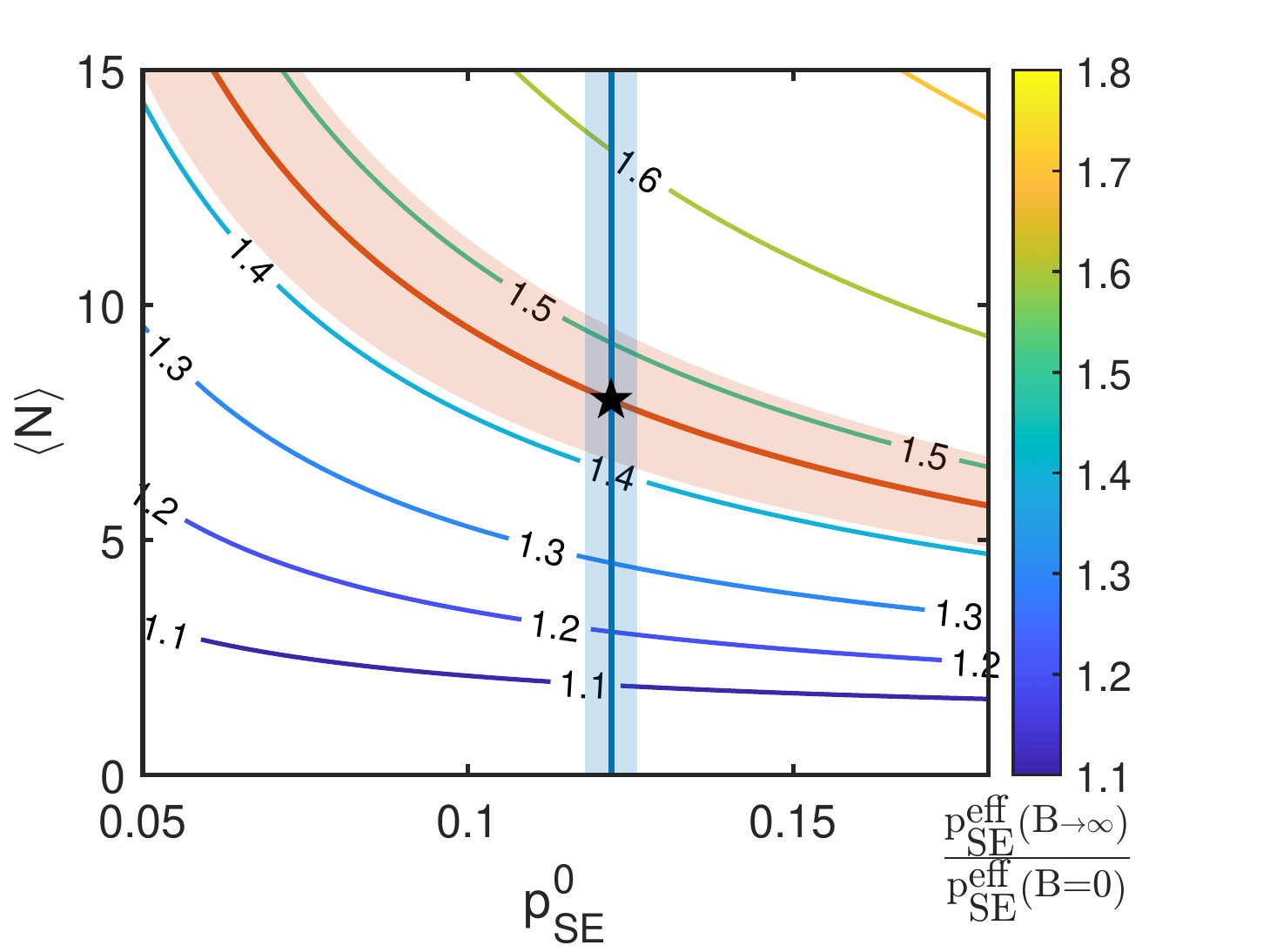}
    \caption{\textbf{Estimating the lifetime of bound state from the simple model.} Countour lines are exothermic spin-exchange amplification, $p_{SE}^\textrm{eff}(B\to \infty)/p^\textrm{eff}_{SE}(B=0)$, calculated by the simple bound state model, for different mean number of short-range collisions, $\langle N \rangle$, and short-range spin-exchange probability, $p_\textrm{SE}^0$.
    Red and blue bold lines are the measured ratios and short-range spin-exchange probability, respectively, with 1$\sigma$ confidence bound in shaded area. The star is indicating the mean number of short-range collisions, $\langle N \rangle ^{exp}=8(2)$.}
    \label{fig:find_N_exp}
\end{figure}
\stepcounter{extended_data_fig}
If there is no magnetic field, the short-range spin-exchange events are independent. Therefore, for a given $n$, the number of spin-exchange events, $n_{\textrm{SE}}$, has a binomial distribution. In the experiment, only odd number of spin-exchange events are detected, and therefore the effective measured spin-exchange probability is given by
\begin{equation}
    p^{\textrm{eff}}_{\textrm{SE}}(B=0)=\sum_{n=1}^\infty (1-p)^{n-1} p \sum_{n_{SE}\ odd}^{n} B(n_{\textrm{SE}};n,p^0_{\textrm{SE}}),
     \label{eq:p_SE_eff_B_0}
\end{equation} where 
\begin{equation}
    B(n_{\textrm{SE}};n,p^0_{\textrm{SE}})=\binom{n}{n_{\textrm{SE}}}(p^0_{\textrm{SE}})^{n_{\textrm{SE}}}(1-p^0_{\textrm{SE}})^{n-n_{\textrm{SE}}},
\end{equation}
is the binomial distribution.
On the other hand, if the magnetic field is larger than the binding energy, only one exothermic spin-exchange event is allowed. In this case the effective probability is
\begin{equation}
     p^{\textrm{eff}}_{\textrm{SE}}(B\to\infty)=\sum_{n=1}^\infty (1-p)^{n-1} p \sum_{m=1}^{n}(1-p^0_{\textrm{SE}}))^{m-1}p^0_{\textrm{SE}}
     \label{eq:p_SE_eff_B_high}
\end{equation}

Then, we can calculate the the ratio $ p^{\textrm{eff}}_{\textrm{SE}}(B=0)/ p^{\textrm{eff}}_{\textrm{SE}}(B\to \infty)$ for any N and p, as shown in \ref{fig:find_N_exp}. From the probability of exothermic spin-exchange as a function of the magnetic field (in Fig.~\ref{fig:SE_B}) we can find the ratio $p_{\textrm{SE}}^{\textrm{eff}}(B=0)/p_{\textrm{SE}}^{\textrm{eff}}(B\to \infty)=1.46(5)$, as shown in the red shaded region in \ref{fig:find_N_exp}. On the other hand, the short-range spin-exchange probability was found in the fit to the exothermic data in Fig.~\ref{fig:SE_EMM}, denoted by the blue line. From crossing these two curves, we can estimate the mean number of short-range collisions, $\langle N \rangle^\textrm{exp} = 8(2)$.

\section*{Short-range collision frequency during bound state formation}
We estimate the time between subsequent collisions by solving the one-dimensional motion in the relative frame-of-reference. The Hamiltonian of the relative motion for a free atom-ion pair is
\begin{equation}
   H=\frac{1}{2}\mu\dot{r}^2-\frac{C_4}{2r^4},
\end{equation}
where $r$ is the relative atom-ion separation, $\mu$ is the reduced mass and $C_4$ is the constant of the interaction. Assume that the initial energy of the system is negative $-E_0=-\frac{C_4}{2r_{\textrm{max}}^4}$, i.e. the particles are bound with a maximal distance $R$. This energy is a constant of motion and we can express the velocity as a function of the position

\begin{equation}
    \dot{r}=\frac{dr}{dt}=\sqrt{\frac{1}{\mu}}\sqrt{\frac{C_4}{r^4}-2E_0}.
\end{equation}

By integrating this equation, we can express the time of falling from a distance $r_{\textrm{max}}$ to the origin by
\begin{equation}
t=\sqrt{\frac{\mu}{C_4}}\int_{r_{\textrm{max}}}^0\frac{r^2dr}{\sqrt{1-\frac{2E_0r^4}{C_4}}}.
\end{equation}

The integral is analytically solvable and gives

\begin{equation}
    t=\sqrt{\frac{\mu}{9C_4}}\frac{\sqrt{\pi}\Gamma(\frac{7}{4})}{\Gamma(\frac{5}{4})}r_{\textrm{max}}^3\approx 
 1.8\sqrt{\frac{\mu}{9C_4}}r_{\textrm{max}}^3,
\end{equation}
where $\Gamma(x)$ is the Gamma function.

\section*{Center-of-mass and relative motion coupling}
In this section, we analyze the coupling between center-of-mass (COM) and relative motion for a simplified version of the Hamiltonian in Eq.~\ref{eq:Hamiltonian}, considering only one dimension and time independent fields (i.e.~setting $q=0$). The simplified Hamiltonian is given by
\begin{equation}\label{eq:H_1d_simple}
        H =\frac{p_\textrm{i}^2}{2m_\textrm{i}}+\frac{m_\textrm{i}a\Omega^2_{RF}}{8} r_{\textrm{i}}^2+       \frac{p_\textrm{a}^2}{2m_\textrm{a}}+V(r).
\end{equation}
We can represent this Hamiltonian using the relative frame with coordinate $r=r_\textrm{i}-r_\textrm{a}$ and conjugate momentum $p$, and COM frame with coordinate, $R=(m_\textrm{i} r_\textrm{i}+m_\textrm{a}r_\textrm{a})/(m_\textrm{i}+m_\textrm{a})$ and conjugate momentum $P$. In these coordinates, the Hamiltonian in Eq.~(\ref{eq:H_1d_simple}) can be decomposed into
\begin{equation}
    H=H_{\textrm{CM}}(R)+H_{\textrm{rel}}(r)+H_{\textrm{cpl}}(r,R),
\end{equation}
where
    \begin{equation}
\begin{split}
H_{\textrm{CM}}(R) & =\frac{P^{2}}{2M}+\frac{1}{2}m_\textrm{i}\omega^{2}R^{2}\\
H_{\textrm{rel}}(r) & = \frac{p^{2}}{2\mu}+\frac{1}{2}\frac{\mu^{2}}{m_\textrm{i}}\omega^{2}r^{2}+V\left(r\right)\\
H_{\textrm{cpl}}(r,R) & =\omega^{2}\mu R\cdot r.
\end{split}
\label{eq:H_cpl}
\end{equation}
Here $\mu=(m_\textrm{i}m_\textrm{a})/(m_\textrm{i}+m_\textrm{a})$ and $M=m_\textrm{i}+m_\textrm{a}$ are the reduced and total mass, respectively, and $\omega=\sqrt{a}\Omega_{RF}^2/2$ denotes the trapping frequency. The relative motion and the COM motion are coupled by the term $H_\textrm{cpl}$, which enables the bound state formation, depends quadratically on the trap frequency and linearly on the reduced mass. 
We note that in inelastic confinement-induced resonances, associated primarily with pairs of cold neutral atoms, the coupling term is typically described by higher orders of $R\cdot r$ \cite{Sala2012}. Therefore, we expect the this mechanism to be more dominant in atom-ion systems, relative to the case of the neutral atoms.





\end{document}